\documentclass[a4paper,fleqn,usenatbib]{mnras}

\usepackage[T1]{fontenc}
\usepackage{ae,aecompl}

\usepackage{graphicx}	
\usepackage{amsmath}	
\usepackage{amsfonts}
\usepackage{amssymb}	
\usepackage{tabularx}
\usepackage{wasysym}
\usepackage{textcomp}
\usepackage{booktabs}
\usepackage{multirow}

\newcommand*{\revision}[1]{{\color{black}{#1}}}
\newcommand*{\change}[1]{{\color{black}{}}}

\newcommand{\be}{\begin{equation}}
\newcommand{\ee}{\end{equation}}
\newcommand{\bea}{\begin{eqnarray}}
\newcommand{\eea}{\end{eqnarray}}

\newcommand{\km}{\,\hbox{km}}
\newcommand{\mum}{\,\hbox{$\mu$m}}
\newcommand{\cm}{\,\hbox{cm}}

\newcommand{\AU}{\,\hbox{AU}}

\newcommand{\Myr}{\,\hbox{Myr}}

\newcommand{\erg}{\,\hbox{erg}}
\newcommand{\K}{\,\hbox{K}}

\newcommand*{\smin}{s_\mathrm{min}}
\newcommand*{\sblow}{s_\mathrm{blow}}
\newcommand*{\sratio}{s_\mathrm{min}/s_\mathrm{blow}}
\newcommand*{\emax}{\langle{e}\rangle}
\newcommand*{\rdisc}{R_\mathrm{disc}}
\newcommand*{\Tage}{T_\mathrm{age}}
\newcommand*{\tbb}{T_\mathrm{BB}}
\newcommand*{\Rbb}{R_\mathrm{BB}}
\newcommand*{\td}{T_\mathrm{dust}}
\newcommand*{\teff}{T_\mathrm{eff}}
\newcommand*{\Qpr}{Q_\mathrm{pr}}
\newcommand*{\Qabs}{Q_\mathrm{abs}}
\newcommand*{\Qsca}{Q_\mathrm{sca}}

\newcommand*{\me}{M_\oplus}
\newcommand*{\Fd}{f_\mathrm{d}}
\newcommand*{\Ls}{L_{\odot}}
\newcommand*{\Ms}{M_{\odot}}

\newcommand{\squee}{\hspace{-0.05em}}

\title[Dust grain sizes in debris discs]{The \squee dust \squee grain \squee size
\squee -- \squee stellar \squee luminosity \squee trend \squee in \squee debris \squee discs}

\author[N. Pawellek \& A. V. Krivov]{
Nicole Pawellek\thanks{E-mail: nicole.pawellek@uni-jena.de}
and Alexander V. Krivov
\\
Astrophysikalisches Institut und Universit\"atssternwarte,Friedrich-Schiller-Universit\"at Jena, Schillerg\"a\ss{}chen 2-3, 07745 Jena, Germany\\
}

\date{Accepted XXX. Received \today; in original form 29 June 2015}

\pubyear{2015}

\begin{document}
\label{firstpage}
\pagerange{\pageref{firstpage}--\pageref{lastpage}}
\maketitle

\begin{abstract}
The cross section of material in debris discs is thought to be dominated
by the smallest grains
that can still stay in bound orbits despite the repelling action of stellar radiation pressure.
Thus the minimum (and typical) grain size $\smin$ is expected to be close to
the radiation pressure blowout size $\sblow$.
Yet a recent analysis of a sample of Herschel-resolved debris discs
showed the ratio $\sratio$ to systematically decrease with the stellar luminosity
from about ten for solar-type stars to nearly unity in the discs around
the most luminous A-type stars.
Here we explore this trend in more detail, checking how significant it is and seeking to find
possible explanations.
We show that the trend is robust to variation of
the composition and porosity of dust particles.
For any assumed grain properties and stellar parameters, we suggest a recipe of how
to estimate the ``true'' radius of a spatially unresolved debris disc, based solely
on its spectral energy distribution.
The results of our collisional simulations are qualitatively consistent with the trend,
although additional effects may also be at work.
In particular, the lack of grains with small
$\sratio$ for lower luminosity stars might be caused by the
grain surface energy constraint that should limit the size of the smallest
collisional fragments.
Also, a better agreement between the data and the collisional
simulations is achieved when assuming 
debris discs of more luminous stars to have higher dynamical
excitation than those of less luminous primaries.
This would imply that protoplanetary discs of more massive young stars are more efficient in
forming big planetesimals or planets that act as stirrers in the debris discs at the subsequent
evolutionary stage.
\end{abstract}



\section{Introduction}

The size distribution of dust in debris discs is set by 
various processes operating in these systems \citep{krivov-et-al-2000b}.
These include radiation pressure, collisions, transport 
processes, as well as mechanisms that lead to erosion of dust grains 
\citep{wyatt-et-al-2011}.
Of particular importance is the minimum radius $\smin$ of grains.
The dust size distribution in most of the debris discs is steep enough for the particles
with sizes close to $\smin$ to be the most abundant and to carry most of the cross section.
Therefore, they are best visible in scattered light, as well as in
thermal emission in the near- and mid-infrared.
In that sense, the size $\smin$ can also be referred to as the typical size of grains
in debris discs.

In discs around solar- and earlier-type stars, grains smaller than the certain
blowout limit $\sblow$ are expelled by direct radiation pressure. Thus one expects
$\smin$ to be close to $\sblow$. However, collisional simulations have shown
the exact relation between the two to depend on the dynamical excitation of dust-producing
planetesimals in the disc.
The ratio $\sratio$ should be slightly above unity in strongly stirred discs
\citep[e.g.,][]{krivov-et-al-2006,thebault-augereau-2007}, but
much larger than unity for dynamically cold discs
\citep[e.g.,][]{thebault-wu-2008,heng-tremaine-2010,krivov-et-al-2013}. 
There are also several other parameters,
such as the mechanical and optical properties of dust,
the disc radius, and system's age,
which may alter $\smin$, $\sblow$, and their ratio.
Thus deriving the typical dust sizes from observations and comparing them with 
the model predictions allows one to constrain all these disc parameters and to gain deeper insights
into the disc physics.

Since most of the debris discs are detected by their thermal emission,
the easiest way to access the grain sizes is to analyse the temperature retrieved from the spectral energy
distributions (SEDs).
This is because the temperature of different-sized grains at the same distance around the same star
is different. Indeed, dust temperatures have been derived for various 
samples of debris discs, based on the  {\em IRAS}, {\em Spitzer}, and {\em Herschel} data
\citep[e.g.,][]{su-et-al-2006,chen-et-al-2006,rhee-et-al-2007,morales-et-al-2011,morales-et-al-2012,%
eiroa-et-al-2013,chen-et-al-2014,kennedy-wyatt-2014,mittal-et-al-2015}.
These studies uncovered a trend of temperatures increasing towards disc host stars of higher luminosity.
Yet the interpretation of these results in terms of grain sizes was hampered by the degeneracy
between the grain size and disc radius, for most of the discs in the samples being unresolved.

\citet{booth-et-al-2013} were the first to invoke a sample of nine resolved discs around A-type stars
to break this degeneracy.
They found evidence for $\smin$ being close to $\sblow$, as expected from the theory.
Most recently, \citet{pawellek-et-al-2014} did a similar study for a sample of 34 resolved debris discs
over a much broader luminosity range.
They confirmed an increase of the minimum grain size with stellar luminosity, as expected from the fact
the $\sblow$ is larger for more luminous stars.
However, this increase was too flat to be consistent with $\smin \approx \sblow$.
Instead, the $\sratio$ ratio was found to decrease from about ten for solar-type stars
to unity for the most luminous A-type primaries~--- an intriguing effect that needs to be explained.

This paper presents a deeper analysis of the $\sratio$ trend,
shows how it can be used for unresolved discs, and tries to explain it.
We start with an analysis of the trend,
investigating the influence of the grain chemical composition and porosity (section~2)
and splitting the entire sample into subsamples of discs of various dustiness, radii,
\revision{ages, and other parameters}
(section~3).
Implications of the trend for the radius estimates of spatially unresolved debris discs
are considered in section~4.
Two possible explanations for the trend are discussed in section~5
(the surface energy constraint on the size of the collisional fragments)
and section~6 (the possible dependence of the disc stirring level on the stellar luminosity).
Section~7 lists our conclusions.

\section{Analysing the trend: Dust grain properties}

\subsection{Dust compositions}

This paper uses the results of \citet{pawellek-et-al-2014} as a starting point.
The idea was to consider a sample of well-resolved debris discs,
for which the dust location can be measured from the images,
and then to perform an SED fitting to uniquely
infer the minimum grain size $\smin$.
However, \citet{pawellek-et-al-2014} assumed compact, spherical dust grains of pure astrosilicate \citep{draine-2003}.
The question arises, to what degree the results may depend on the dust composition.
To answer it, we re-did the fitting of the entire sample by assuming different dust compositions.

Since we are studying the decrease in the $\sratio$ ratio with stellar luminosity,
it is necessary that the blowout grain size exists.
However, for the two M-stars of the sample of \citet{pawellek-et-al-2014}
this is not the case for any dust composition. Therefore we excluded them from further 
investigations and used only 32 objects of the previous sample.
Another reason to exclude the M-stars was that the presence of the strong stellar wind of 
an unknown strength would introduce one more free parameter, making theoretical predictions for $\smin$
highly uncertain.
The sample used here is given in Table~\ref{tab:stars}.

The SED fitting was done as follows
\citep[for details, the reader is referred to][]{pawellek-et-al-2014}.
Many of the discs in the sample are suspected to have a two-component structure, 
consisting
of the main, cold outer disc (a ``Kuiper belt'') and an additional, warm inner one (an ``asteroid belt'').
Since in this paper we are only concerned with the main, outer component, the warm one
had to be looked for and, if present, subtracted from the SED.
We did that exactly as described in \citet{pawellek-et-al-2014}.
Of the two fitting methods used there, the modified blackbody emission and the size distribution method,
here we only employed the latter one, since it is more reliable and gives a direct handle on the particle sizes.
The fitting itself was done by a simulated annealing algorithm, implemented in the SEDUCE code \citep{mueller-et-al-2009}.
We considered five dust grain models described below.
Altogether, we performed a second component check and a complete fitting of 160 objects (32 for each dust composition),
not counting additional fitting runs~-- e.g. to test extreme grain porosities.

\begin{table*}
 \caption{
 Stellar parameters sorted by stellar luminosity 
 \label{tab:stars}
 }
 \tabcolsep 3pt
    \begin{center}
    \begin{tabular}{rrccrrrrc}
    \toprule
    HD 		& HIP 		& Name 		& SpT 		& $L/L_\odot$ & $T_\text{eff}$ [K] & $M/M_\odot$ & Age [Myr]$^f$ & Age ref \\
    \midrule
    23484	& 17439 	& - 		& K2V 		& 0.41  & 5166  & 0.79 	& 930 	& 1		\\
    104860 	& 58876 	& - 		& F8 		& 1.16  & 5930  & 1.04 	& 200 	& 2		\\
    207129 	& 107649 	& - 		& G2V 		& 1.25  & 5912  & 1.06 	& 2499 	& 1, 3, 4	\\
    10647 	& 7978 		& q1 Eri 	& F9V 		& 1.52  & 6155  & 1.12 	& 1307 	& 1, 4, 5, 6	\\
    48682 	& 32480 	& 56 Aur 	& G0V 		& 1.83  & 6086  & 1.17 	& 1380 	& 1 		\\
    50571 	& 32775 	& HR 2562 	& F5VFe+0.4 	& 3.17  & 6490  & 1.35 	& 449 	& 5, 6, 7	\\
    170773 	& 90936 	& HR 6948 	& F5V 		& 3.44  & 6590  & 1.38 	& 574 	& 5, 7, 8	\\
    218396 	& 114189 	& HR 8799 	& A5V 		& 4.81  & 7380  & 1.51 	& 71 	& 5, 9, 10	\\
    109085 	& 61174 	& $\eta$ Crv 	& F2V 		& 5.00  & 6950  & 1.53 	& 1768 	& 11, 12, 13	\\
    27290 	& 19893 	& $\gamma$ Dor 	& F1V 		& 6.27  & 7070  & 1.62 	& 896 	& 10, 12	\\
    95086 	& 53524 	& - 		& A8III 	& 7.04  & 7530  & 1.70$^b$ 	& 15 	& 7, 14		\\
    195627 	& 101612 	& $\phi 1$ Pav 	& F0V 		& 7.36  & 7200  & 1.69 	& 842 	& 7, 8		\\
    20320 	& 15197 	& $\zeta$ Eri 	& kA4hA9mA9V$^a$& 10.3  & 7575  & 1.85 	& 800 	& 12		\\
    21997 	& 16449 	& HR 1082 	& A3IV/V 	& 11.2  & 8325  & 1.89 	& 44 	& 7, 10, 15	\\
    110411 	& 61960 	& $\varrho$ Vir	& A0V 		& 11.7  & 8710  & 1.91 	& 71 	& 12, 16	\\
    142091 	& 77655 	& $\kappa$ CrB 	& K1IVa 	& 12.5  & 4815  & 1.80$^c$ 	& 2345 	& 17, 18	\\
    102647 	& 57632 	& $\beta$ Leo 	& A3Va 		& 13.2  & 8490  & 1.97 	& 82 	& 12, 16, 19, 20\\
    125162 	& 69732 	& $\lambda$ Boo	& A0p 		& 15.4  & 8550  & 2.05 	& 301 	& 12, 16	\\
    216956 	& 113368 	& Fomalhaut 	& A4V 		& 15.5  & 8195  & 2.06 	& 200 	& 21		\\
    17848 	& 13141 	& $\nu$ Hor 	& A2V 		& 15.7  & 8400  & 2.07 	& 261	& 6, 7, 22	\\
    9672 	& 7345 		& 49 Cet 	& A1V 		& 16.0  & 9000  & 2.07 	& 36 	& 5, 7, 22, 23	\\
    71722 	& 41373 	& HR 3341 	& A0V 		& 18.5  & 8925  & 2.16 	& 100 	& 2 		\\
    182681 	& 95619 	& HR 7380 	& B9V 		& 24.9  & 10000 & 2.33 	& 73 	& 7 		\\
    14055 	& 10670 	& $\gamma$ Tri 	& A1Vnn 	& 25.0  & 9350  & 2.33 	& 160 	& 12		\\
    161868 	& 87108 	& $\gamma$ Oph 	& A0V 		& 26.0  & 9020  & 2.36 	& 276 	& 7, 10, 16	\\
    188228 	& 98495 	& $\epsilon$ Pav& A0Va 		& 26.6  & 10190 & 2.37 	& 50 	& 12, 16	\\
    10939 	& 8241 		& q2 Eri  	& A1V 		& 31.3  & 9200  & 2.47 	& 352 	& 7, 8, 24	\\
    71155 	& 41307 	& 30 Mon 	& A0V 		& 35.7  & 9770  & 2.56 	& 169 	& 12, 16 	\\
    172167 	& 91262 	& Vega 		& A0V 		& 51.8  & 9530  & 2.83 	& 265 	& 16, 25	\\
    139006 	& 76267 	& $\alpha$ CrB 	& A0V 		& 57.7  & 9220  & 3.50$^d$ 	& 291 	& 12, 16 	\\
    95418 	& 53910 	& $\beta$ UMa 	& A1IVps 	& 58.2  & 9130  & 2.70$^e$ 	& 305 	& 12, 16 	\\
    13161 	& 10064 	& $\beta$ Tri 	& A5III 	& 73.8  & 8010  & 4.90$^d$ 	& 730 	& 12		\\
    \bottomrule
    \end{tabular}%
    \end{center}
    
\noindent
{\em Notes:}\\[0mm]
The effective temperatures and ages 
are averaged over the listed literature values.\\
$^a$Gray-Corbally notation. See App.~A2 in \citet{trilling-et-al-2007}
for its explanation. 
The majority of the stellar masses was computed from the luminosities by means of a standard relation 
$M\propto L^{1/3.8}$ for main sequence stars. Exceptions for other luminosity classes or close binaries are the following:
$^b$stellar mass from \cite{moor-et-al-2013} (giant);
$^c$stellar mass from \cite{bonsor-et-al-2013b} (subgiant);
$^d$sum of the stellar masses from \cite{kennedy-et-al-2012b} (close binaries);
$^e$stellar mass from \cite{booth-et-al-2013} (subgiant).
$^f$For each star with more than one age reference, the age given is the geometric mean of the values reported in those papers.

\medskip
\noindent
{\em Age references:}\\[0mm]
[1]~\cite{eiroa-et-al-2013};
[2]~\cite{morales-et-al-2013};
[3]~\cite{loehne-et-al-2011};
[4]~\cite{trilling-et-al-2008};
[5]~\cite{moor-et-al-2006};
[6]~\cite{rhee-et-al-2007};
[7]~\cite{moor-et-al-2014};
[8]~\cite{chen-et-al-2014};
[9]~\cite{marois-et-al-2010};
[10]~\cite{chen-et-al-2006};
[11]~\cite{duchene-et-al-2014};
[12]~\cite{vican-et-al-2012};
[13]~\cite{beichman-et-al-2006};
[14]~\cite{moor-et-al-2013};
[15]~\cite{moor-et-al-2011};
[16]~\cite{su-et-al-2006};
[17]~\cite{bonsor-et-al-2013b};
[18]~\cite{bonsor-et-al-2014};
[19]~\cite{churcher-et-al-2011};
[20]~\cite{song-et-al-2001};
[21]~\cite{acke-et-al-2012};
[22]~\cite{nielsen-et-al-2013};
[23]~\cite{roberge-et-al-2013};
[24]~\cite{morales-et-al-2011};
[25]~\cite{sibthorpe-et-al-2010}.

\end{table*}%

Specifically, we selected the compositions listed in Table~\ref{tab:dust} (percentages are volume fractions).
We used the \citet{draine-2003} optical constants for astrosilicate,  the \citet{zubko-et-al-1996} data for carbon 
(their ``ACAR sample'') and the \citet{li-greenberg-1998} data for ice particles. 
For pure astrosilicate, vacuum inclusions were added to simulate possible porosity.
The Bruggeman mixing rule \citep{bohren-huffman-1983} was applied to compute the refractive 
indices of mixtures, and the Mie theory was employed to calculate the efficiencies.

\begin{table}
\caption{Dust compositions
\label{tab:dust}
}
\tabcolsep 3pt
  \begin{center}
  \begin{tabular}{lc}
  \toprule
  Dust composition& $\varrho$ [g/cm$^3$]\\
  \midrule
  50\% astrosilicate + 50\% vacuum        & 1.65\\
  50\% astrosilicate + 50\% ice           & 2.25\\
  100\% astrosilicate                     & 3.30\\
  50\% astrosilicate + 50\% carbon        & 2.63\\
  100\% \revision{carbon}                 & 1.95\\
  \bottomrule
  \end{tabular}
  
\end{center}

\end{table}

\begin{figure}
   \centering
   \includegraphics[width=0.5\textwidth,angle=0]{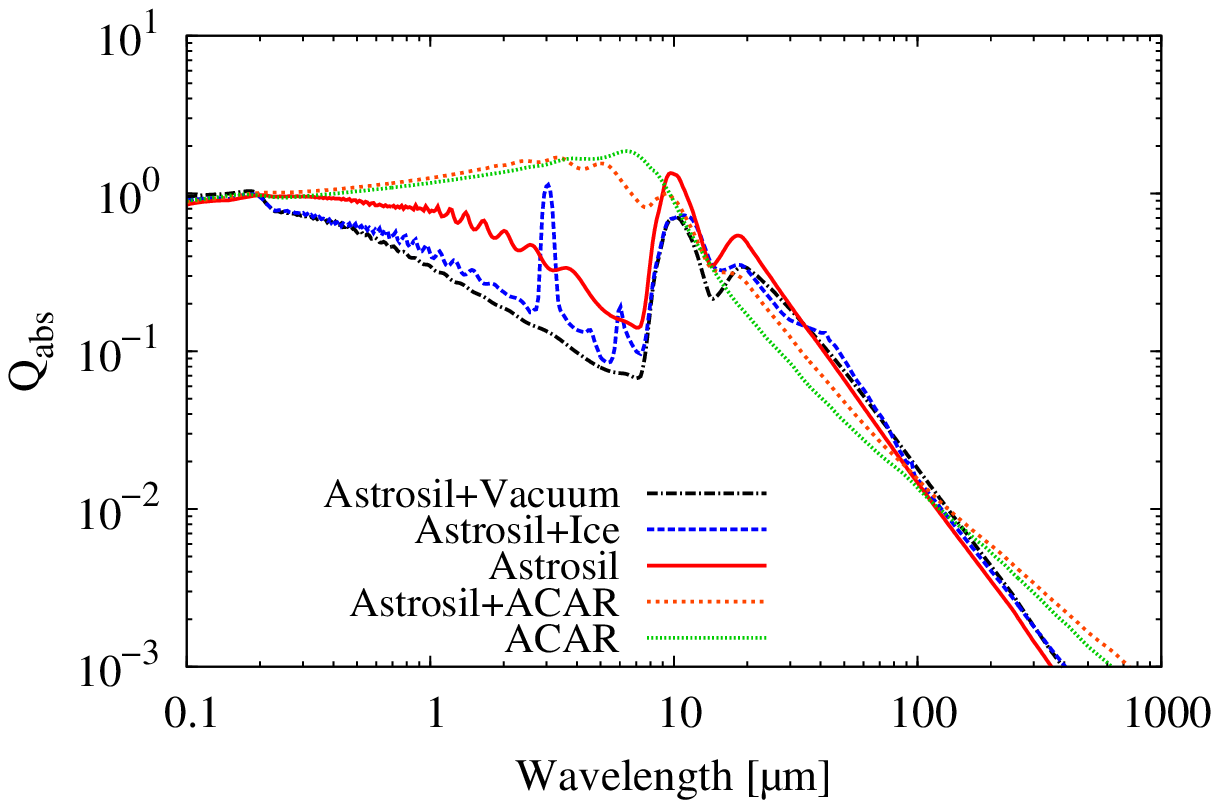}
   \vspace*{-5mm}
   \caption{Absorption efficiency $\Qabs (\lambda)$ for $s = 1\mum$.
   \label{fig:Qabs}
   }
   \end{figure}

\begin{figure}
   \centering
   \includegraphics[width=0.5\textwidth,angle=0]{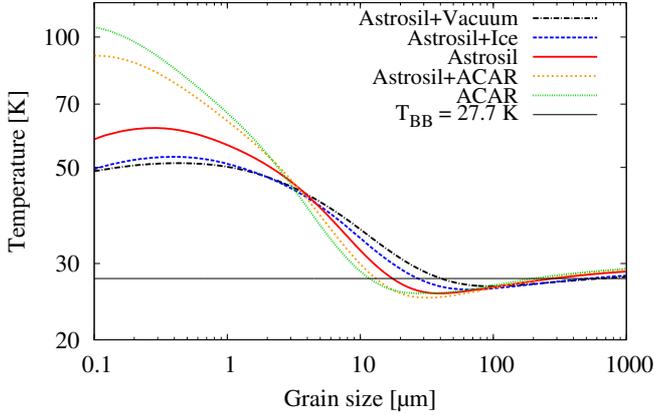}
   \vspace*{-5mm}
   \caption{
   Dust grain temperature $\td (s)$ for a star with $1\Ls$ at a radius of $100\AU$. 
   $\tbb$ is the blackbody temperature.
   \label{fig:Tdust}
   }
  \end{figure}

The dust compositions in Table~\ref{tab:dust} are ordered by the increasing grain temperature $\td$ for a grain size of 
$1\mum$ at a distance of $100\AU$ from a Sun-like star.
For all five compositions, Fig.~\ref{fig:Qabs} plots the absorption efficiency $\Qabs$ of 
$1\mum$-sized grains as a function of wavelength.
Figure \ref{fig:Tdust} presents the dust temperature $\td$ around a star of
solar luminosity as a function of grain radius $s$.
The temperature curves for all materials are similar in shape.
Small grains ($s<10\mum$) are warmer than blackbody,
whereas the temperature of grains with a size larger than $10\mum$ is close to the blackbody temperature.
The grains with sizes between $0.1\mum$ and $1\mum$ (depending on the dust composition)
are the hottest.
Both smaller and larger grains are colder \citep{krivov-et-al-2006}.

\subsection{Calculation of the blowout size}

The ratio of radiation pressure and gravitational force
is given by \citep{burns-et-al-1979}
\be
 \beta = \frac{F_\mathrm{rad}}{F_\mathrm{grav}} = \frac{3L}{16\pi GMc}\frac{\Qpr}{\varrho s} .
\label{beta}
\ee
Here, $G$ is the gravitational constant, $L$ the stellar luminosity, $M$ the stellar mass, $c$ the speed of light, 
$\Qpr$ the radiation pressure efficiency averaged over the stellar spectrum, $\varrho$ the density of the grains and 
$s$ the grain radius.
The grain size that corresponds to $\beta = 0.5$ 
is the blowout grain size $\sblow$.
\citet{pawellek-et-al-2014} set $\Qpr$ to one (geometric optics approximation).
To determine $\sblow$ more accurately, we calculated $\Qpr$ for each of the material compositions
as \citep{burns-et-al-1979}
\be
 \Qpr \equiv \Qabs + \Qsca(1-\langle\cos(\vartheta)\rangle) .
\ee
Here,
$\Qsca$ is the scattering efficiency
and $\langle\cos(\vartheta)\rangle$
the anisotropy parameter with the scattering angle $\vartheta$:
\be
 \langle\cos(\vartheta)\rangle \equiv \int\limits_{4\pi} f(\vartheta)\cos(\vartheta)d\Omega ,
\ee
where $\Omega$ is the solid angle and $f(\vartheta)$ the phase function 
that we computed with Mie theory \citep{bohren-huffman-1983}.
The averaging of $\Qabs$ and $\Qsca$ over the stellar spectra was done with the aid of
the PHOENIX/GAIA model grid \citep{brott-hauschildt-2005} and,
for two stars with $\teff \ge 10000\K$, ATLAS9 models
\citep{castelli-kurucz-2004}.

\begin{figure}
   \centering
   \includegraphics[width=0.50\textwidth,angle=0]{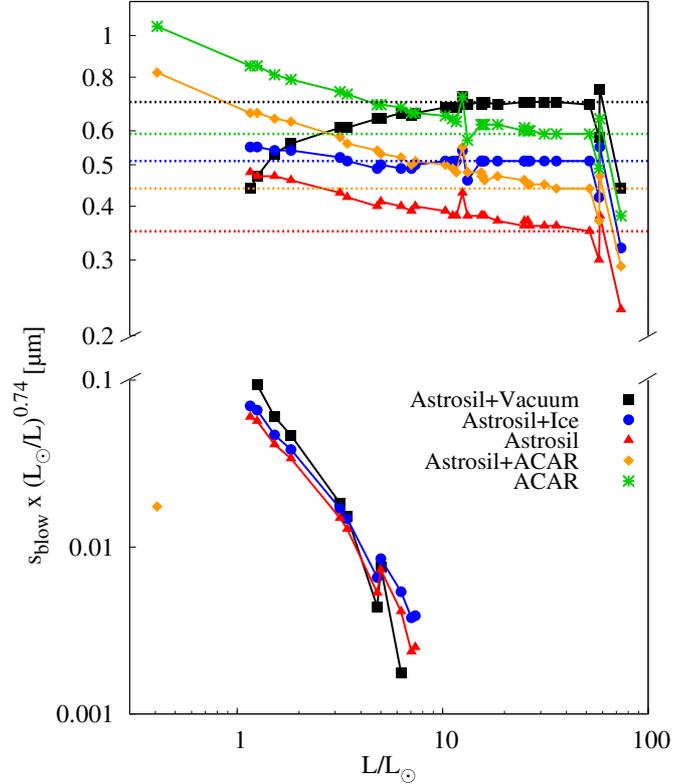}
   \vspace*{-10mm}
   \caption{
   Blowout grain size, multiplied with $(M/\Ms)/(L/\Ls)$, as a function of stellar 
   luminosity. Different colours correspond to different dust compositions, as indicated in the legend.
   Symbols with connecting solid lines: results for size-dependent $\Qpr$ obtained with the Mie theory
   and for actual stellar masses from Table~\ref{tab:stars};
   horizontal dashed lines: results for $\Qpr=1$ and assuming the main-sequence mass-luminosity relation.
   The vertical offset between the symbols and lines that represent different compositions is caused 
   by their different bulk densities, see Table~\ref{tab:dust}.
   For stars with $\Ls \la L \la 10\Ls$, except for pure carbon, the second blowout limit 
   appears in the bottom 
   part of the plot. Grains in blowout orbits are those between the upper and lower branches.
   For the lowest-luminosity star in the sample (HD~23484 with $L=0.41 \Ls$), a single or double blowout limit only 
   exists for two out of five mixtures.
   The strongest outliers amongst the symbols
   are stars with masses departing from the main-sequence mass-luminosity
   relation. These are the A1-subgiant $\beta$~UMa with $L/\Ls = 58.2$ and the close binaries $\alpha$~CrB ($L/\Ls = 57.7$)
   and the A1III-star $\beta$~Tri ($L/\Ls = 73.8$).
   Some other apparent outliers
   reflect the fact that $\sblow$ for any individual star depends on $\Qabs$ and $\Qsca$
   averaged over the stellar photospheric spectrum of that particular star.
   For example, $\kappa$~CrB with $L/\Ls = 12.5$ is a K1-subgiant with a temperature
   twice lower than that of its neighbouring stars in the figure. 
   This results in the larger averaged values of $\Qabs$, $\Qsca$, and $\sblow$ compared to its neighbours.
   For still other stars and particular grain compositions, the stellar spectrum peaks at the maxima
   of the resonant oscillations of the Mie-calculated $\Qabs$ and $\Qsca$, which also makes $\sblow$
   different from that of the adjacent stars.
   This is particularly the case for $\beta$~Leo with $L/\Ls = 13.2$, if
   the astrosilicate-ice and astrosilicate-carbon mixtures are assumed.
   \label{fig:sblow}
   }
   \end{figure}

Having found $\Qpr (s)$, expression (\ref{beta}) should be equated to 0.5, yielding an equation for $\sblow$.
That equation
has one root for luminous stars.
For stars of moderate luminosities, it may have two solutions,
in which case the grains with sizes between the two roots are in unbound orbits.
For the least luminous stars, the equation may not have solutions at all, meaning that
$\beta$ is always smaller than $0.5$ and no $\sblow$ exists.
We solved this equation numerically.
The results for all five compositions and for all stars in our sample are shown in 
Fig.~\ref{fig:sblow} with symbols.
More exactly, we plot the product $\sblow (\Ls/L)^{0.74}$ to eliminate the trend
$\sblow \propto L/M \propto L^{-0.74}$ 
(assuming here $M \propto L^{1/3.8}$ as appropriate for main-sequence stars).
This makes the results for different compositions more easily distinguishable.
For comparison, with dashed lines we depict the results calculated 
in the geometric optics approximation
(e.g., with $\Qpr = 1$)
{and assuming again $M \propto L^{1/3.8}$ for all stars};
multiplication by $(\Ls/L)^{0.74}$ renders these lines horizontal.

\subsection{Results}

   \begin{figure*}
   \centering
   \includegraphics[width=0.49\textwidth,angle=0]{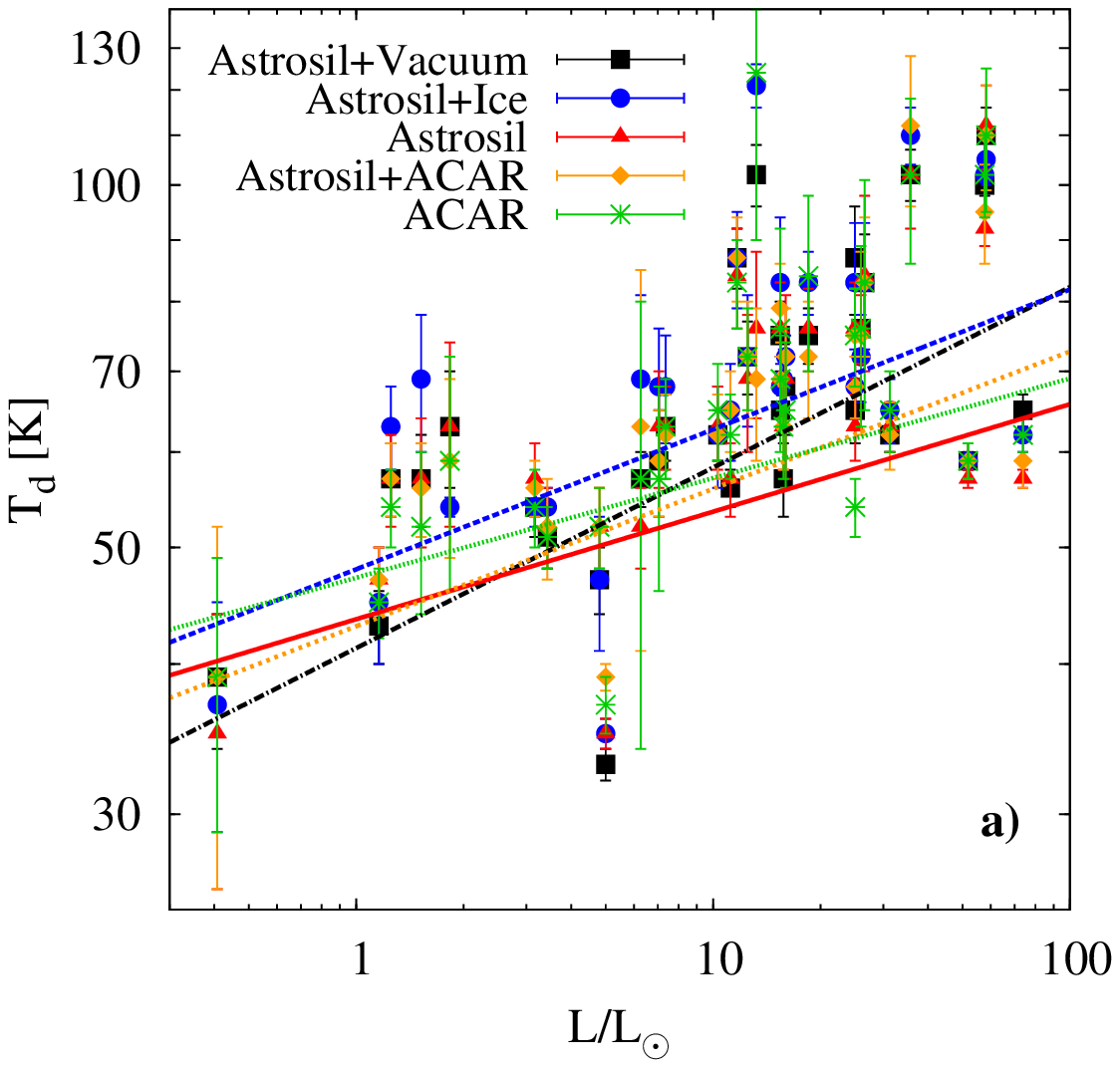}
   \includegraphics[width=0.49\textwidth,angle=0]{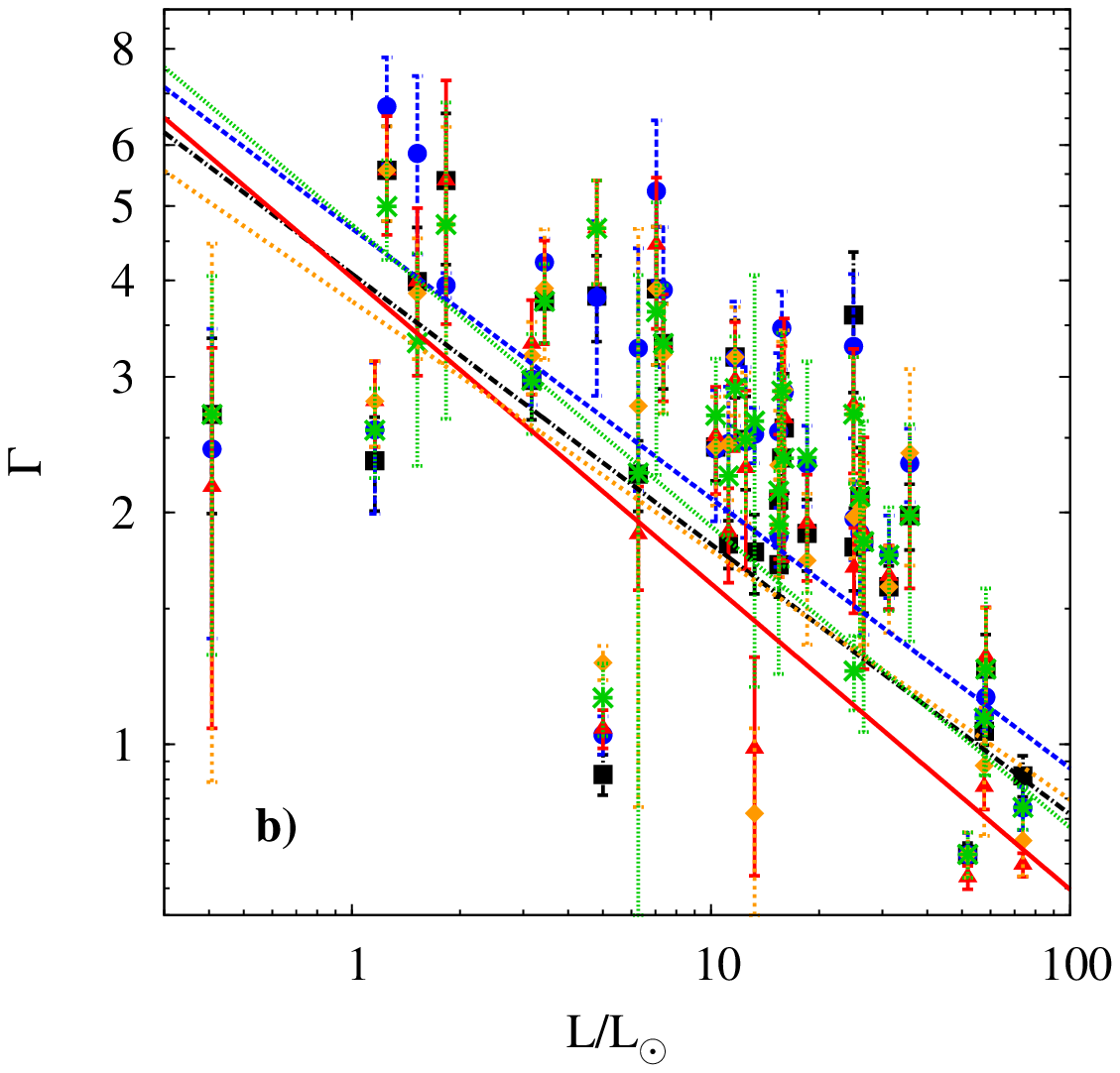}\\[-5mm]
   \includegraphics[width=0.49\textwidth,angle=0]{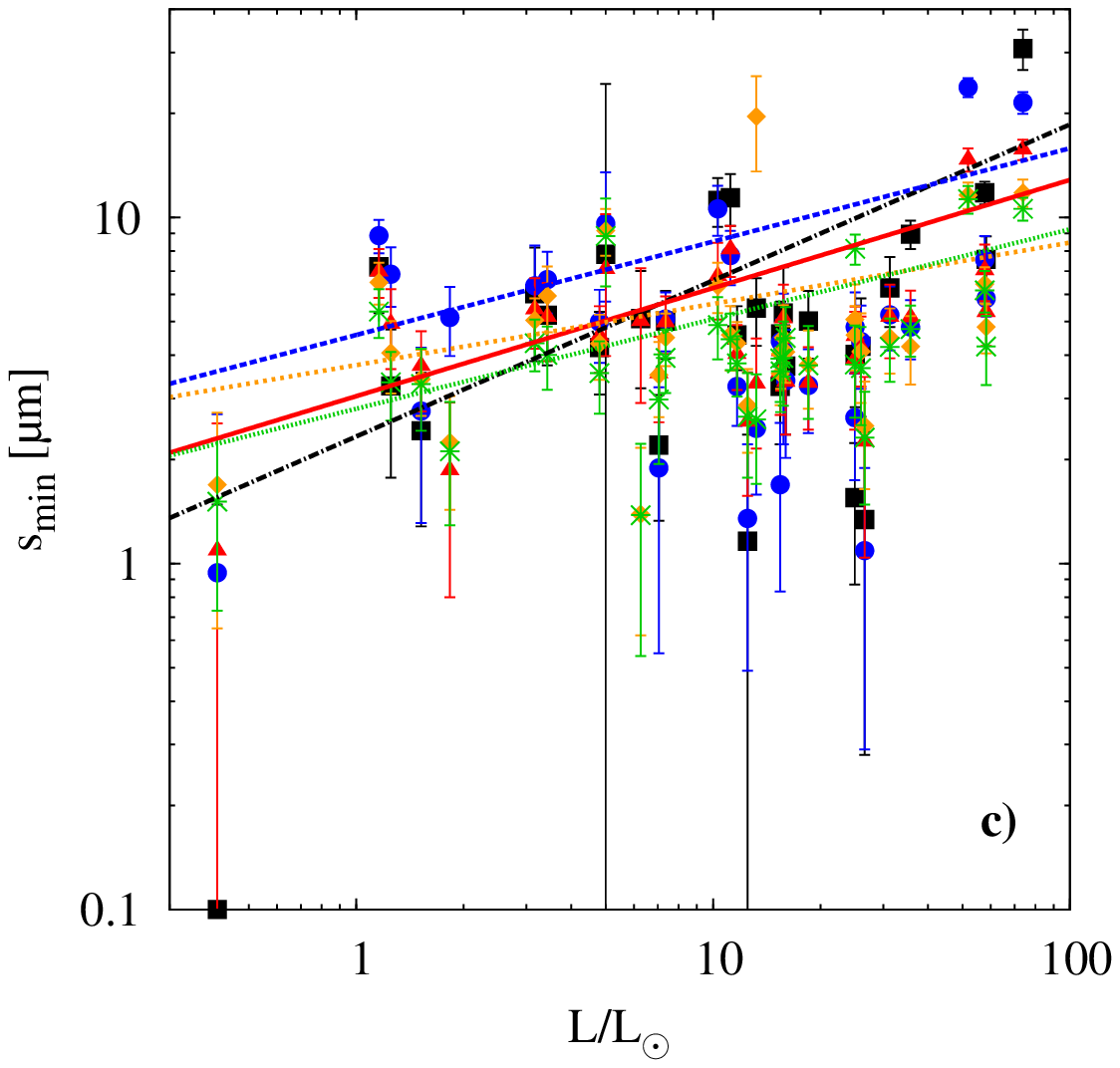}
   \includegraphics[width=0.49\textwidth,angle=0]{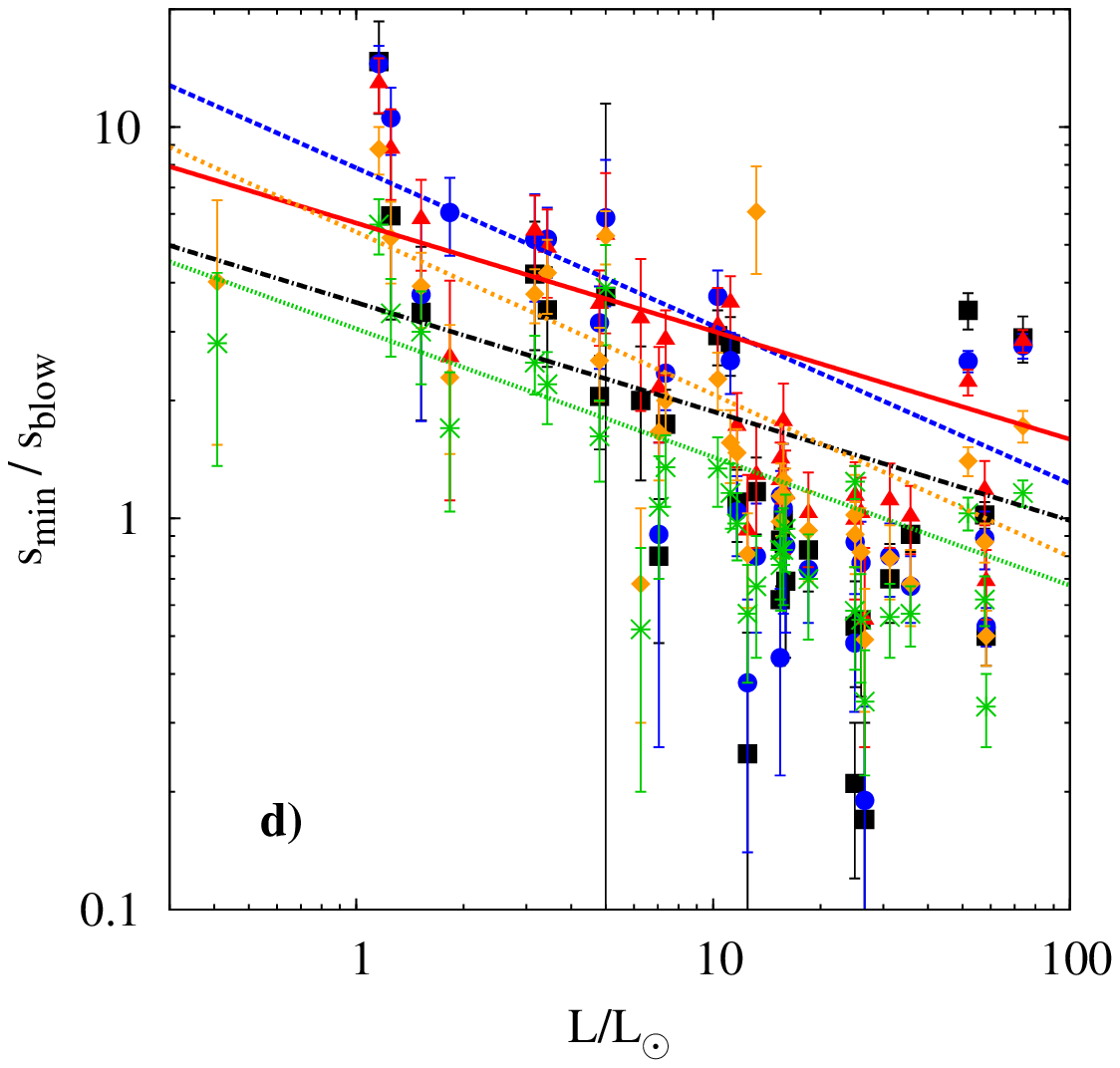}
   \vspace*{-5mm}
   \caption{
   Various dust and disc parameters as functions of stellar luminosity for different dust compositions:
            (a) dust temperature,
            (b) disc's true radius to its blackbody radius, $\Gamma$,
            (c) minimum grain size $\smin$,
            (d) grain size ratio $\sratio$.
            Similar to the previous figures, different colours denote different dust compositions. 
            Symbols with error bars are fitting results for individual discs in our sample.
            A straight line of a certain colour is a best-fit trend line through the symbols of the same colour.
            For two objects (HD~48682 and HD~27290) and for two mixtures (astrosilicate+vacuum and astrosilicate+ice),
            the best-fit $\smin$ and $\sratio$ are too small for the plotting range of the panels c) and d).
   \label{fig:parm_L}
   }
   \end{figure*}
   
Figure~\ref{fig:parm_L} depicts the SED fitting results for all five compositions, presenting
the dust temperature $\td$,
the ratio of the true disc radius to the blackbody radius $\Gamma$,
the grain size $\smin$,
and the ratio $\sratio$. 
It demonstrates that all these parameters reveal more or less clear trends with the stellar luminosity:
the dust gets warmer, the disc radius goes down to the blackbody value,
the typical grain size increases, but the typical grain size in the blowout units decreases.
All these results are qualitatively the same
for different dust compositions and are consistent with those obtained previously
for the pure astrosilicate \citep{pawellek-et-al-2014}.
   \begin{figure*}
   \centering
   \includegraphics[width=0.48\textwidth,angle=0]{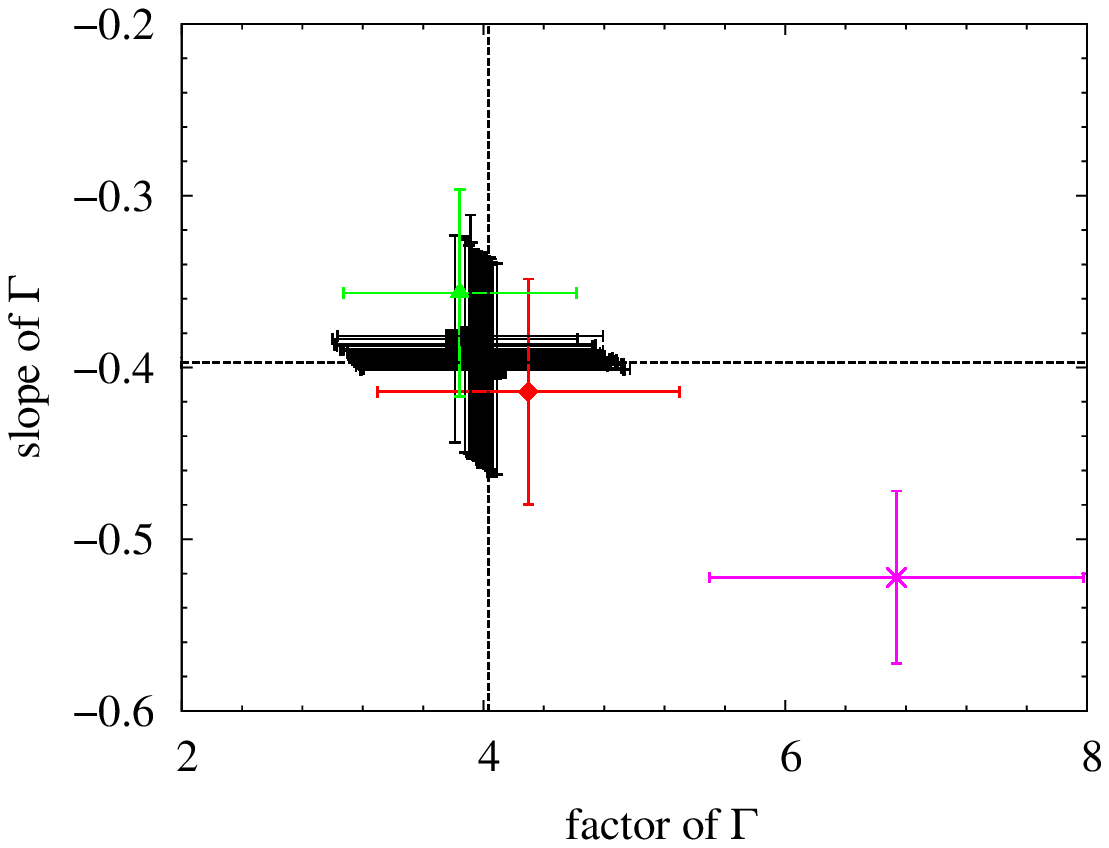}
   \includegraphics[width=0.48\textwidth,angle=0]{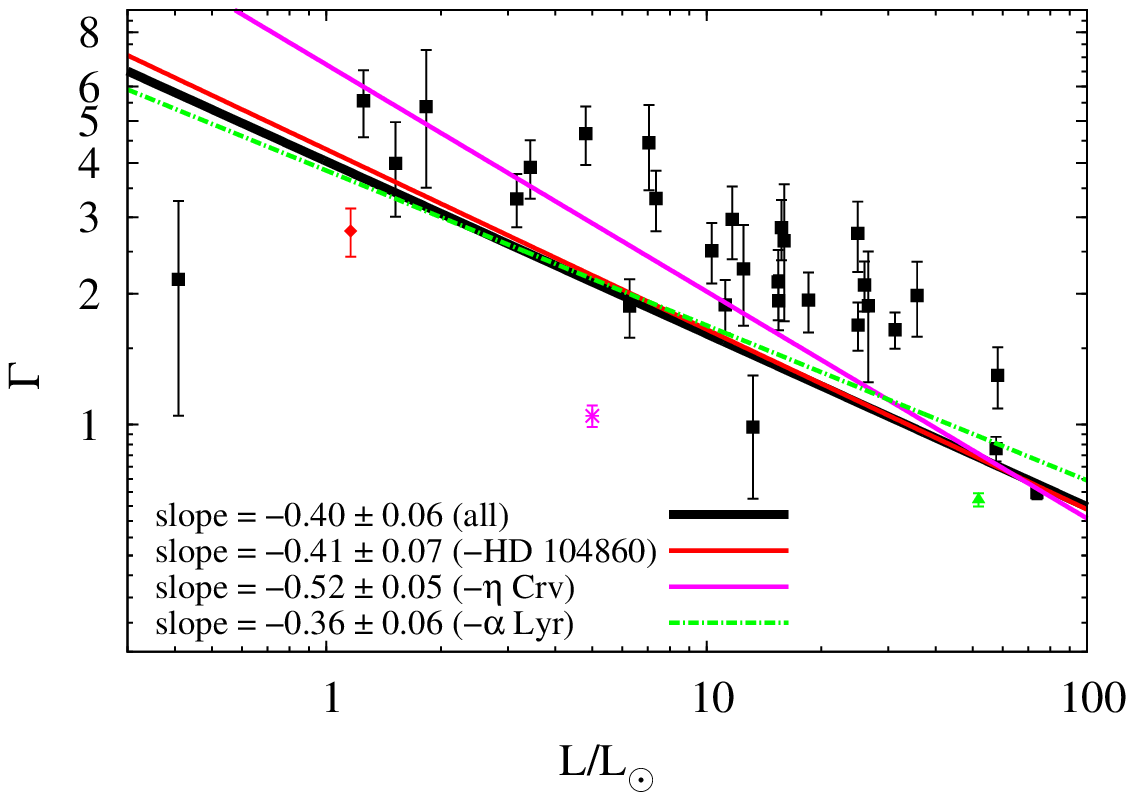}\\
   \includegraphics[width=0.48\textwidth,angle=0]{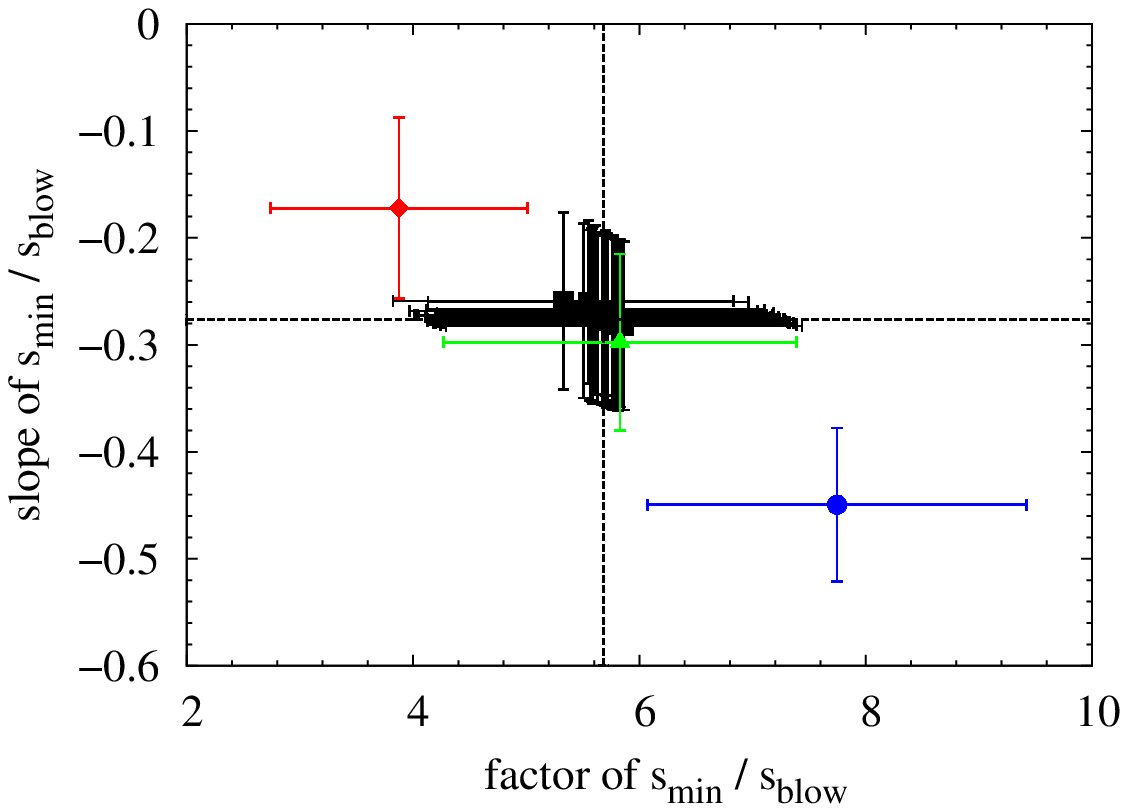}
   \includegraphics[width=0.48\textwidth,angle=0]{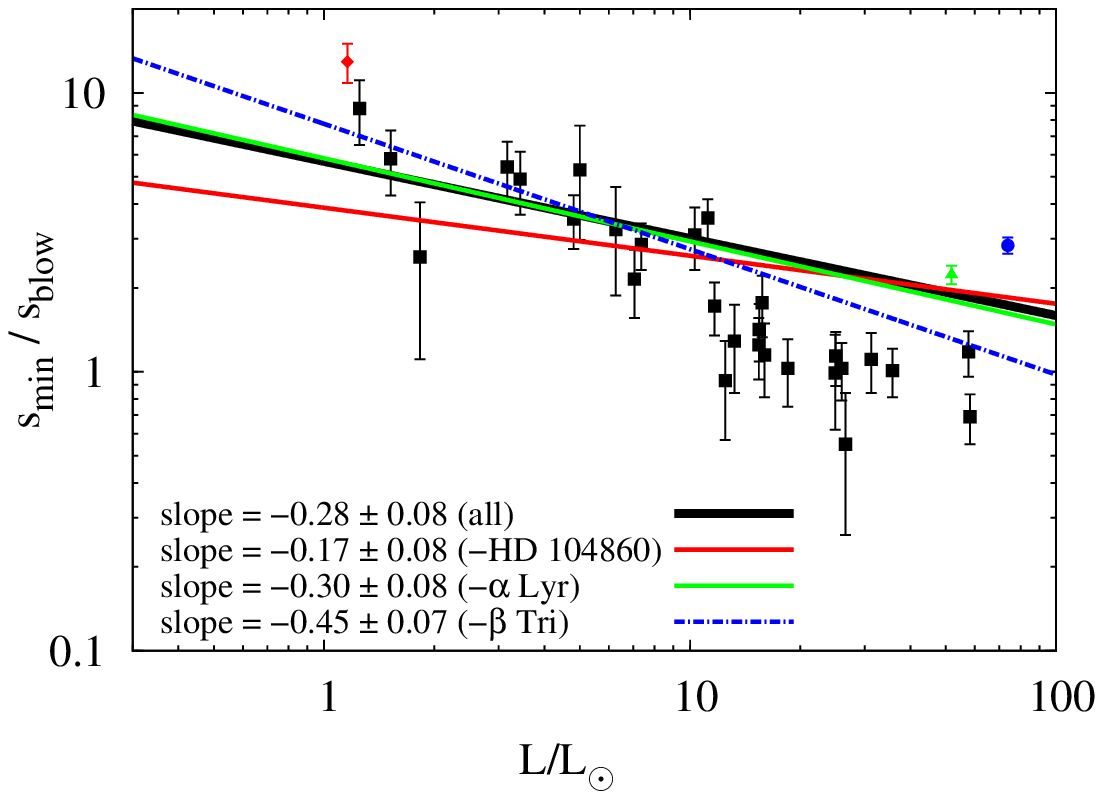}\\
   \caption{
    Identification of outliers in $\Gamma(L)$ (top panels) and $\sratio(L)$ (bottom panels)
    for pure astrosilicate.
    Left: the factors $A$ and slopes $B$ of the trend lines in the form $A \; \left( L/\Ls \right)^B$,
    calculated by excluding from the sample one data point at a time.
    Error bars are uncertainties in $A$ and $B$ returned by the fitting.
    The ``cross-hairs'' depict the best fit through the whole sample.
    The strongest outliers are shown in colour: red is HD~104860,
    magenta is $\eta$~Crv,
    green is Vega,
    and blue is $\beta$~Tri.
    Right: $\Gamma$ and size ratio versus stellar luminosity for pure astrosilicate 
    with and without outliers. The colour coding is the same as on the left.
    The black line gives the trend line for the entire sample.
    The line of a certain colour is the best fit without the object of the same colour.
   \label{fig:outliers}
   }
   \end{figure*}   

Nevertheless, quantitative differences are apparent.
In particular, for porous particles, 
the $\sratio$ dependence on stellar luminosity is flatter than for compact ones.
This is easy to understand.
Figure~\ref{fig:Tdust} shows that for $s < 3\mum$ the temperature of porous grains is the
lowest of all compositions used and conversely, porous grains with sizes larger than $3\mum$ 
are hotter than the other mixtures.
As a result, for early-type stars the temperature of a porous astrosilicate grain
of size $\smin$ is higher than that of a compact astrosilicate particle of the same size.
To reproduce the SED of a disc around such a star, the porous grains need to be 
larger than the compact ones.
Conversely, for late-type stars a porous grain with size $\smin$
is colder than a compact one and therefore the porous particles must be 
smaller than the compact ones.
The result is a steeper increase of the minimum grain size of porous grains with the stellar luminosity,
and thus a gentler decrease of the $\sratio$ ratio, compared 
to the other dust compositions.

Still, for the assumed porosity of 50\%, the trend is not fully erased.
Therefore, we tried to find the minimum degree of porosity that would make
the $\sratio$ ratio independent \revision{of} the stellar luminosity.
It turned out that for most of the discs the quality of the SED fits, 
measured by $\chi^2$, gets poorer with increasing degree of porosity.
For example, the disc of HD~50571
yields the reduced $\chi^2 = 1.85$ for a porosity of zero
(for pure astrosilicate).
With a 50\% porosity, $\chi^2$ increases to 1.86, and with 90\% to 4.55. 

We conclude that the decrease of $\sratio$ towards higher stellar luminosities
identified by \citet{pawellek-et-al-2014} for compact astrosilicate particles
is pretty robust with respect to variation of the assumed grain properties.

\section{Analysing the trend: Outliers and subsamples}
\label{sec:subsamples}

\subsection{Outliers}


There is a concern about the effect of outliers in
the plots of derived parameters versus stellar luminosity.
For example, $\eta$~Crv is the apparent outlier with a low disc
temperature and $\Gamma$ in Fig.~\ref{fig:parm_L}a,b, and is
largely the reason the best-fit lines are below most of the data.
Also, two of the four most luminous stars having rather large derived $\smin$
and small associated uncertainties
(Vega and $\beta$~Tri) might affect the trend lines
of $\smin$ and $\sratio$ in Fig.~\ref{fig:parm_L}c,d.
Thus some investigation of how much individual
systems are influencing the results is needed to check the
robustness of the conclusion of the $\sratio$ trend.
 
A formal way to identify the outliers and quantify their effect is as follows.
We go over the whole sample,
remove one data point by one, each time calculating the best-fit lines, and look
for those discs \revision{whose removal alters}
the regression line the most strongly.
We do this procedure for two plots that play the major role in the rest of the paper:
$\Gamma(L)$ and $\sratio(L)$.

Figure~\ref{fig:outliers} depicts the results for pure astrosilicate.
In terms of $\Gamma$, the strongest outliers are
$\eta$~Crv, HD~104860, and Vega.
Removing $\eta$~Crv changes the slope of the trend line more strongly (by 0.12)
than the uncertainty returned from the best-fit to the whole population (0.06)!
For $\sratio$, three strongest outliers are
$\beta$~Tri, HD~104860, and Vega. 
This confirms that a few individual outliers influence the best-fit lines appreciably.
On the other hand, dropping each of the objects other than those listed above
has a minor effect on the best-fit lines.

For these reasons,  we chose to exclude three respective strongest outliers
from the sample for the rest of the paper.
In sections 3 and 5, where $\sratio$ is analysed, we remove
$\beta$~Tri, HD~104860, and Vega. Besides, automatically excluded is the lowest-luminosity
star HD~23484, for which~--- in the case of astrosilicate~--- no blowout limit exists and
thus $\sratio$ is undefined. This implies a set of 28 stars.
In a similar style, in section 4, dealing the $\Gamma$-ratio and disc radii,
we discard $\eta$~Crv, HD~104860, and Vega and work with a set of remaining 29 stars.

``Refining'' the sample in such a way is done in order
not to bias the results with a few systems,
which are the strongest outliers and for which the uncertainties are atypicially small.
In doing so, we also keep in mind possible peculiarities of the discarded stars
that may explain them being outliers.
For instance, HD~104860 is amongst the most poorly resolved sources in the sample,
which makes its radius less certain.
Besides, the low luminosity of this star makes us believe that the dust can be efficiently
transported inward from the parent belt, implying that the disc radius measured from the image
may be smaller than the true radius of the parent belt.
Correcting for this would return a smaller $\smin$ than derived here, reducing the deviation
from the $\sratio$-trend line.
The strongest $\Gamma$-outlier, $\eta$~Crv, is known to be unusual in many other respects
\citep[see, e.g.,][]{duchene-et-al-2014}.
In the case of Vega, the assumed luminosity of $52\Ls$ may be in error.
This star is a rapid rotator \citep{peterson-et-al-2006,aufdenberg-et-al-2006},
which makes stellar parameters functions of the stellar latitude.
The ``equatorial stellar luminosity'' seen by the dust
could be as small as $28\Ls$ and may be significantly lower than the
``polar luminosity'' of $57\Ls$ measured from the Earth
\citep[see, e.g.,][for a discussion]{mueller-et-al-2009}.
Shifting the Vega point to the left in Figure~\ref{fig:outliers} could make it a non-outlier.
As far as $\beta$~Tri is concerned, this is the most luminous star in the sample.
It has some peculiarities, too, being one of only two giants in the sample and a close binary and,
as such, may not be representative of the whole population.

\subsection{Extracting and comparing subsamples}

The grain sizes and the size ratios may
also depend on physical parameters of the systems other than the stellar luminosity
(for instance, on disc's fractional luminosity, disc radius,
or system's age). One suitable method to prove this is as follows.
One can split the entire sample 
of now 28 objects (i.e., without HD~23484, HD~104860, Vega, and $\beta$~Tri)
into a pair of subsamples with $n_1$ and $n_2$ objects
($n_1+n_2 = n = 28$) according to the physical parameter $P$ selected.
The two subsamples could consist, for instance, of systems with $P$ smaller and greater
than the parameter's median value $P_\mathrm{med}$, respectively.
In that case, $n_1 = n_2 = 14$. Then, one takes $\sratio (L)$ for the two subsamples
separately and derives the best-fit log-log trend lines with the slopes $b_1 \pm SE(b_1)$
and $b_2 \pm SE(b_2)$, where $SE(x)$  is a standard error of $x$. The final step is to
find out whether the relations $\sratio (L)$ in the two subsamples are statistically
significantly different. (The null hypothesis is obviously that they are not.)
To this end, one computes the Student's t-score as
\be
  t = |b_1 -b_2| / SE(b_1-b_2)
  \label{t-score}
\ee
with
\be
  SE(b_1 -b_2) = \sqrt{SE(b_1)^2+SE(b_2)^2}
  \label{SE}
\ee
and calculates the probability $p$ that the null hypothesis is true on 
$n_1+n_2-4 = 24$
degrees of freedom.
We use a strict criterion and consider two subsamples significally different if
the two-tailed $p<0.05$.

\begin{figure}
   \centering
   \includegraphics[width=0.5\textwidth,angle=0]{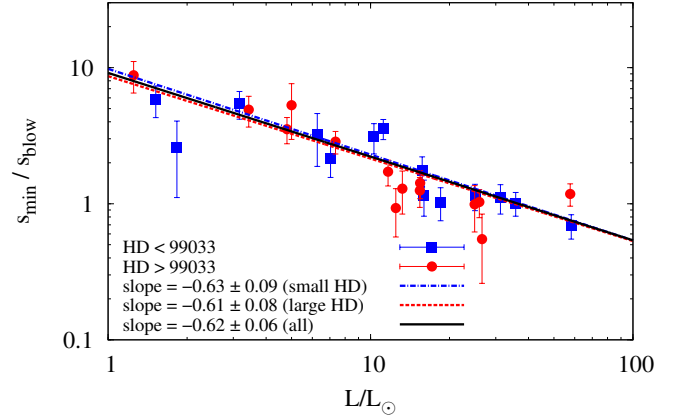}
   \vspace*{-5mm}
   \caption{
    Size ratio vs. stellar luminosity for pure astrosilicate, for two subsamples
    based on the HD numbers.
    Blue line and squares: HD number smaller than the median value 99033; 
    red line and circles: HD number larger than that;
    black line: the entire sample.
   \label{fig:HD}
   }
   \end{figure}

Since, with its 28
objects, our sample is not particularly large, a question arises if
this procedure is feasible. To test this, we chose as $P$ the HD number of the debris disc
stars. This parameter is unphysical. Since there are no reasons to believe that whatever
properties of debris discs should depend on the right assension of their primaries, our procedure
should confirm the validity of the null hypothesis. The results are visualised
in Fig.~\ref{fig:HD}, while the statistical parameters of the regression comparison are listed
in Table~\ref{tab:statist}. The probability is {$p=0.87$}, 
so that the two subsamples are
statistically indistinguishable.

\begin{table}
  \caption{
	    Comparing the regressions for $\sratio$ in pairs of 
	    subsamples for $P<P_\mathrm{med}$ and $P>P_\mathrm{med}$
	    \label{tab:statist}
	  }
  \tabcolsep 3pt
  \begin{center}
    \begin{tabular}{l|ccclc}
      \toprule
	$P$			& $b_1 \pm SE(b_1)$	& $b_2 \pm SE(b_2)$	& t-score	& $p$		& Verdict \\
      \midrule
	HD			& $-0.63 \pm 0.09$ 	& $-0.61 \pm 0.08$	& $0.17$	&$0.87$		& same \\
	$\Fd$			& $-0.48 \pm 0.11$ 	& $-0.67 \pm 0.08$	& $1.40$	&$0.17$		& same \\
	$\rdisc$		& $-0.53 \pm 0.08$ 	& $-0.69 \pm 0.09$	& $1.33$	&$0.20$		& same \\
	$\Tage$			& $-0.74 \pm 0.15$ 	& $-0.58 \pm 0.06$	& $0.99$	&$0.33$		& same \\
	$F_{100}$		& $-0.60 \pm 0.07$ 	& $-0.65 \pm 0.10$	& $0.41$	&$0.69$		& same \\
	$F_{100}/\text{Extent}$	& $-0.70 \pm 0.08$ 	& $-0.57 \pm 0.09$	& $1.08$	&$0.29$		& same \\
	FWHM			& $-0.51 \pm 0.15$ 	& $-0.65 \pm 0.06$	& $0.87$	&$0.40$		& same \\
      \bottomrule
    \end{tabular}
  \end{center}
\end{table}

\subsection{Discs of low and high fractional luminosity}

\begin{figure}
   \centering
   \includegraphics[width=0.5\textwidth,angle=0]{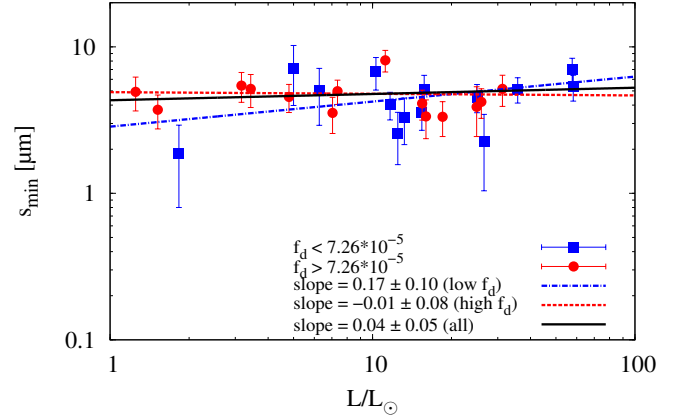}\\
   \includegraphics[width=0.5\textwidth,angle=0]{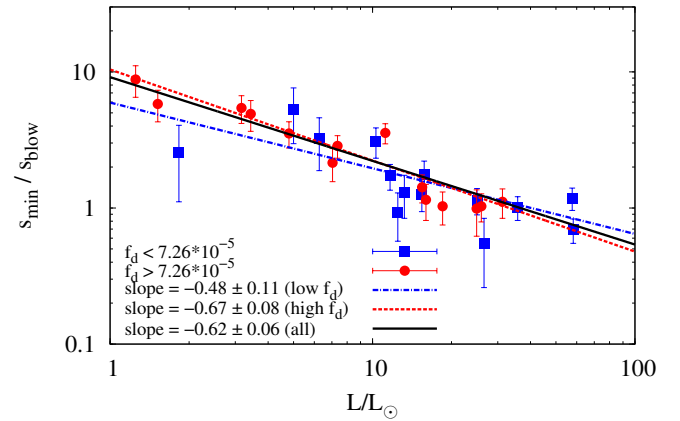}
   \vspace*{-5mm}
   \caption{
    Minimum size (top) and size ratio (bottom) vs. stellar luminosity, assuming pure astrosilicate,
    for the subsamples of low and high fractional luminosity discs.
    Blue line and squares: $\Fd < 7.26 \times 10^{-5}$, 
    red line and circles: $\Fd > 7.26 \times 10^{-5}$, 
    black line: the entire sample.
    \change{Sentence deleted}
    \label{fig:fd}
   }
   \end{figure}

We first take two subsamples that comprise discs with fractional luminosity
lower and higher than the median value $7.26\times 10^{-5}$.
Figure \ref{fig:fd} shows both the minimum 
grain size and the ratio of the minimum to the blowout size in both groups
as a function of stellar luminosity for the reference material, astrosilicate.
Clearly, the host stars of the discs of both classes provide a broad coverage
of stellar luminosities: from $1.8\Ls$ to $58\Ls$ and
from $1.3\Ls$ to $31\Ls$, respectively.
For high-$\Fd$ discs, the minimum size $\smin$ turns out to be $\approx 5\mum$,
nearly independent of the stellar luminosity
(the slope, or regression coefficient, being $-0.01 \pm 0.08$ only).
However, $\smin$ of low-$\Fd$ discs, having the slope
of $0.17 \pm 0.10$), is also $2\sigma$-consistent with being constant.
The $\sratio$ ratio decreases 
with stellar luminosity,
but this trend is slightly stronger for high fractional luminosity discs 
(regression coefficient of $-0.67 \pm 0.08$)
than for low fractional luminosity ones
($-0.48 \pm 0.11$).
Besides, the high fractional luminosity discs also reveal less scatter
(Peason's correlation coefficient of
$r=-0.95_{-0.03}^{+0.11}$, Spearman's 
$r_s=-0.94_{-0.06}^{+0.22}$)
than those with low $\Fd$
($r=-0.75_{-0.16}^{+0.39}$, 
$r_s=-0.80_{-0.15}^{+0.35}$).

Table~\ref{tab:statist} presents the formal comparison of the two subsamples (for
$\sratio$ ratio only; but we checked that for $\smin$ the results are nearly identical).
The $p$-probability is $0.17$, so that the subsamples do not differ significantly.

   \begin{figure}
   \centering
   \includegraphics[width=0.5\textwidth,angle=0]{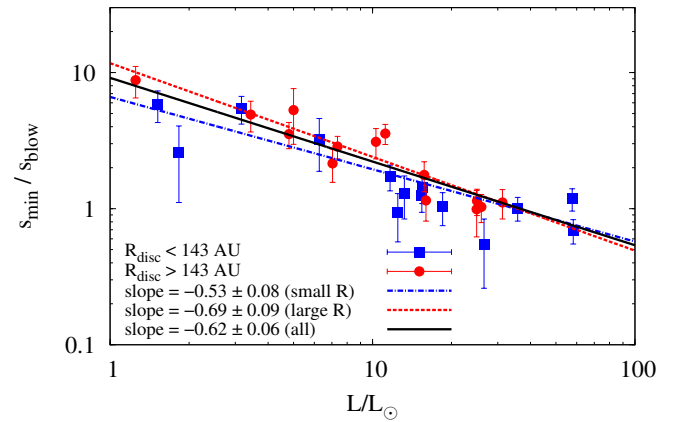}
   \vspace*{-5mm}
   \caption{
    Same as Figs.~\ref{fig:HD} and \ref{fig:fd} bottom,
    but for the two subsamples
    based on the disc radius.
    Blue line and squares: $\rdisc < 143\AU$;
    red line and circles:  $\rdisc > 143\AU$; 
    black line: the entire sample.
   \label{fig:R}
   }
   \end{figure}

   \begin{figure}
   \centering
   \includegraphics[width=0.5\textwidth,angle=0]{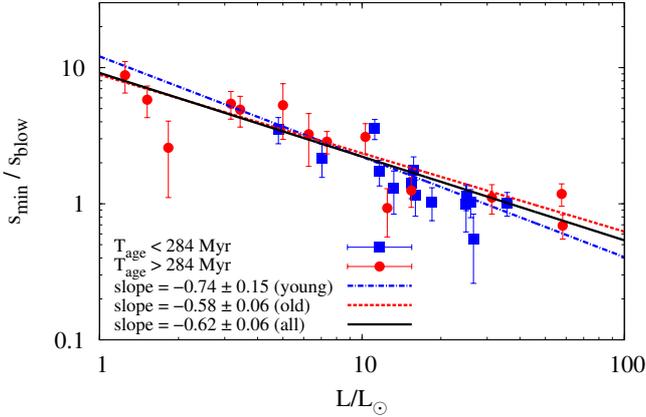}
   \vspace*{-5mm}
   \caption{
    Same as Figs.~\ref{fig:HD}--\ref{fig:R}, but for the two subsamples
    based on system's age.
    Blue line and squares: systems younger than $284\Myr$;
    red line and circles:  systems older than $284\Myr$;
    black line: the entire sample.
   \label{fig:t}
   }
   \end{figure}

Even though the differences between the subsamples are of low significance,
they are marginally visible.
There are two possibilities to explain them.
One is that a higher $\Fd$ implies a more pronounced infrared excess.
The SEDs of such discs can therefore be fitted in a more reliable way and the results would be less uncertain.
This view is supported by the fact that the scatter of the data for high-$\Fd$ discs around the trend line
is smaller
(i.e., the correlation between the size or size ratio and the luminosity is stronger).
Indeed, the standard error of the slope for the high-$\Fd$ discs (0.08) 
is smaller than that
of the low-$\Fd$ ones (0.11).

Another possibility is of physical nature: it can be that discs with high and low fractional luminosity
are physically different in one or another respect. If this is true, this might be reflected by
a systematic difference in one or another key parameter, such as for instance
the disc radius or system's age, between the two groups of discs.

\subsection{Small vs large discs and young vs old discs}

To prove whether the discs with higher and lower dust luminosity are 
physically different,
we have checked the disc radius $\rdisc$ and the age of the systems $\Tage$ 
in both groups of discs.
It turned out that the discs with $\Fd > 7.26~\times~10^{-5}$
are on the average larger 
($\rdisc = 170\AU \pm 49\AU$) and younger 
{($\Tage = 488\Myr \pm 685\Myr$)
than those with  $\Fd < 7.26~\times~10^{-5}$ 
($\rdisc = 120\AU \pm 47\AU$ and 
${\Tage = 634\Myr \pm 720\Myr}$). 
Since the standard deviations of the radii and ages are quite large, 
this conclusion has to be tested statistically.
The null hypothesis is
that the disc radii and ages in the two subsamples are indistinguishable.

As long as the radii and ages are not distributed normally, we choose the 
$k$-sample
Anderson--Darling (AD) test \citep{scholz-stephens-1987}
which is independent of a specific distribution and the sample size and is 
sufficiently sensitive to the tails of the distribution.
We assume that the disc radii have an accuracy of 7\% \citep{pawellek-et-al-2014} and
conservatively estimate the age determination accuracy as 
$25\% \times \log_{10}(\Tage/\Myr)$.
The AD test provides the testing value $A^2$, which is given by
\be
  A^2 = \frac{1}{n_1+n_2}\sum\limits_{i=1}^{k}\frac{1}{n_i}\sum\limits_{j=1}^{n_1+n_2-1}\frac{((n_1+n_2) M_{ij} - jn_i)^2}{j(n_1+n_2-j)} .
\ee
Here, $n_i = 14$ is the size of the $i$-th sample, $k=2$ 
the number of samples and $M_{ij}$ the number of 
data points in the $i$-th sample smaller than the data point $Z_j$ of the
pooled ordered sample ($Z_1<...< Z_{n_1+n_2}$).
The $A^2$ parameter is standardised to 
\be
  T\equiv\frac{A^2-(k-1)}{\sigma},
\ee
where $\sigma$ is the standard deviation of $A^2$.
The null hypothesis 
is rejected at a significance level $\alpha$ if $T \geq t_{k-1}(\alpha)$.
%
%
For $k = 2$ and a significance level of ${\alpha=0.05}$, the critical 
percentile value is $t_1 = 1.96$.
For the disc radii we find 
${T=3.56\pm0.79}$ and for the ages 
$T=-0.38 \pm 0.84$.
Thus the radii of the discs with high and low $\Fd$ may
be statistically different, whereas their ages are clearly not.
   
Despite this result, we made two additional tests.
In one of them we divided the whole sample into the subsamples
of small and large disc radii, separated by the
median radius of $143\AU$ (Fig.~\ref{fig:R}).
In the second test,
we took the median age of the sample ($284\Myr$) and 
divided the sample in young ($\Tage < 284\Myr$) 
and old ($\Tage > 284\Myr$) objects (Fig.~\ref{fig:t}).
The statistical parameters for the two-sample tests 
(see Eqs.~\ref{t-score}--\ref{SE}) are listed in
Table~\ref{tab:statist}. The $p$-probabilities that the radius and age 
subsamples are the same
are $0.20$ and $0.33$,  
respectively.
This supports the view that marginal differences between the discs of higher 
and lower fractional luminosity
are more likely related to the difference in quality of the fits rather 
than are caused by physical reasons.

\subsection{Faint vs bright discs}


Another way to split the sample that may be illuminating
is by the absolute integrated flux from the disc, or perhaps that flux divided by the disc extent.
This would be a test for whether the results are being significantly affected by a
systematic related to the images, since for example discs brighter in
absolute terms may be derived to be larger simply because more of the disc
can be seen above the noise.
To prove this, we now split the sample by the discs'
absolute {\em integrated} brightness at 100~$\mu$m (Fig. \ref{fig:F100})
and by their absolute {\em surface} brightness (Fig. \ref{fig:F100_pro_extent}).
Both faint and bright discs, in terms of integrated flux and surface flux,
are found around stars of all luminosities.
Judging by trend lines, we find no significant differences between these subsamples.
The $p$-probabilities for these tests are $0.69$ and $0.29$, respectively
(Table~\ref{tab:statist}).

   \begin{figure}
   \centering
   \includegraphics[width=0.5\textwidth,angle=0]{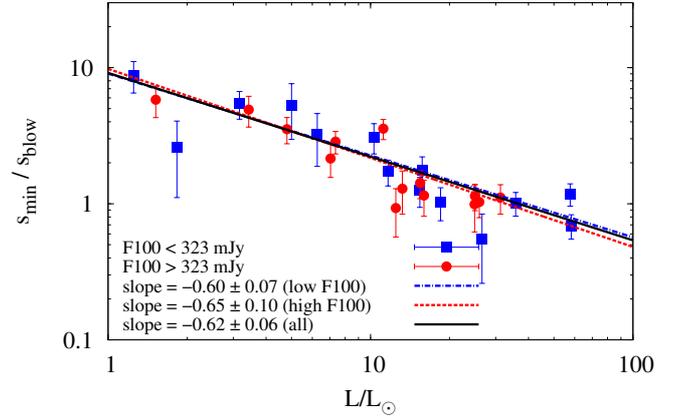}
   \vspace*{-5mm}
   \caption{
    Same as Figs.~\ref{fig:HD}--\ref{fig:t},
    but for the two subsamples
    based on disc's absolute integrated brightness.
    Blue line and squares: faint discs;
    red line and circles:  bright discs;
    black line: the entire sample.
   \label{fig:F100}
   }
   \end{figure}

   \begin{figure}
   \centering
   \includegraphics[width=0.5\textwidth,angle=0]{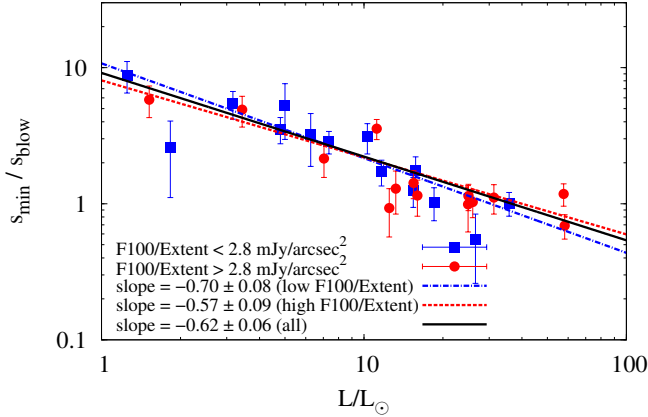}
   \vspace*{-5mm}
   \caption{
    Same as Figs.~\ref{fig:HD}--\ref{fig:F100},
    but for the two subsamples
    based on disc's absolute {\em surface} brightness.
    Blue line and squares: faint discs;
    red line and circles:  bright discs;
    black line: the entire sample.
   \label{fig:F100_pro_extent}
   }
   \end{figure}

\subsection{Marginally-resolved vs well-resolved discs}


One more possibility is that trend lines are affected by the 
uncertainties with which
the disc radii have been measured. This can be characterised by the 
discs' full width at half maximum (FWHM) at $100\mum$
which can be compared to the {\em Herschel}/PACS point-spread function (PSF) width at the same 
wavelength, $6.8\arcsec$. 
We split the sample into 14 discs with $\textrm{FWHM} < 10.9\arcsec$ 
(i.e., $\textrm{FWHM} < 1.6 \textrm{PSF}$)
and another 14 discs with $\textrm{FWHM} > 10.9\arcsec$ 
(i.e., $\textrm{FWHM} > 1.6 \textrm{PSF}$).
The results are depicted in Fig.~\ref{fig:FWHM}.
Blue points indicating marginally-resolved discs are mostly found in the right portion of the plot
and conversely, well-resolved discs concentrate in the left part of the figure.
This reveals that FWHM has an anti-correlation with the stellar luminosity.
As discussed in \citet{pawellek-et-al-2014}, this reason is that more luminous stars in the sample
are more distant on the average, so that their discs often have a smaller angular size and
are more poorly resolved.
However, the trend lines in both subsamples are pretty much the same
($p=0.40$, see Table~\ref{tab:statist}).

   \begin{figure}
   \centering
   \includegraphics[width=0.5\textwidth,angle=0]{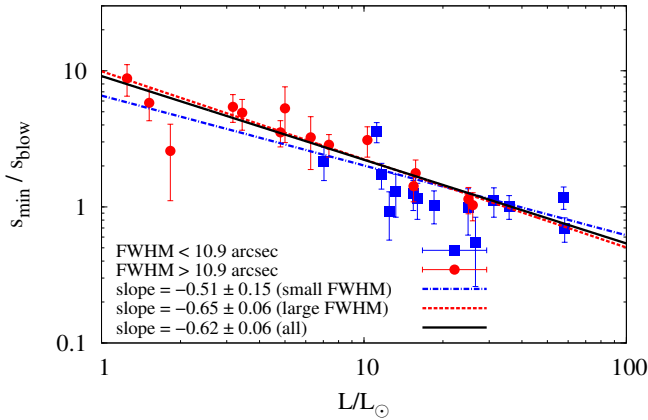}
   \vspace*{-5mm}
   \caption{
    Same as Figs.~\ref{fig:HD}--\ref{fig:F100_pro_extent},
    but for the two subsamples
    based on disc's FWHM at $100\mum$.
    Blue line and squares: marginally resolved discs;
    red line and circles:  well-resolved discs;
    black line: the entire sample.
   \label{fig:FWHM}
   }
   \end{figure}

\section{Using the trend: Application to unresolved debris discs}

We now return to the ratio of the disc radius to the blackbody radius, $\Gamma$.
Its dependence on the stellar luminosity
was presented in Fig.~\ref{fig:parm_L}b
for the complete sample of 32 discs and different dust compositions.
In view of the analysis presented in section~\ref{sec:subsamples}, we now remove
the three strongest outliers, namely $\eta$~Crv, HD~104860, and Vega.
The best-fit relations between $\Gamma$ and the luminosity of the central star have the form
\be
 \Gamma = A \; \left( L/\Ls \right)^B,
\label{Gamma}
\ee
where the power-law coefficients $A$ and $B$ for the refined sample of 29 stars and
all five material compositions are listed in Table~\ref{tab:coeff}.

Table~\ref{tab:coeff} shows that the trend lines of $\Gamma(L)$ are not very dissimilar for
all of the mixtures except for pure astrosilicate. In this case, the trend is the steepest and
clearly predicts disc radii that are too large for stars of solar and subsolar luminosity.
This might be an argument against using pure astrosilicate as a ``reference material'' for debris dust.
Although often used, it is worth reminding that astrosilicate does not exist in nature.
It is an artificial material that was originally designed to describe the
observations in the interstellar medium \citep{draine-lee-1984}, and there is no justification
to apply it to circumstellar discs, either protoplanetary or debris ones.
This conclusion seems also consistent with numerous studies of individual debris discs that showed the pure
astrosilicate to fit the SEDs more poorly than dust compositions combining silicates with ice, carbon,
and \revision{vacuum} \citep[e.g.,][among many others]
{lebreton-et-al-2012,donaldson-et-al-2013,morales-et-al-2013,schueppler-et-al-2014}.

   \begin{figure}
   \centering
   \includegraphics[width=0.5\textwidth,angle=0]{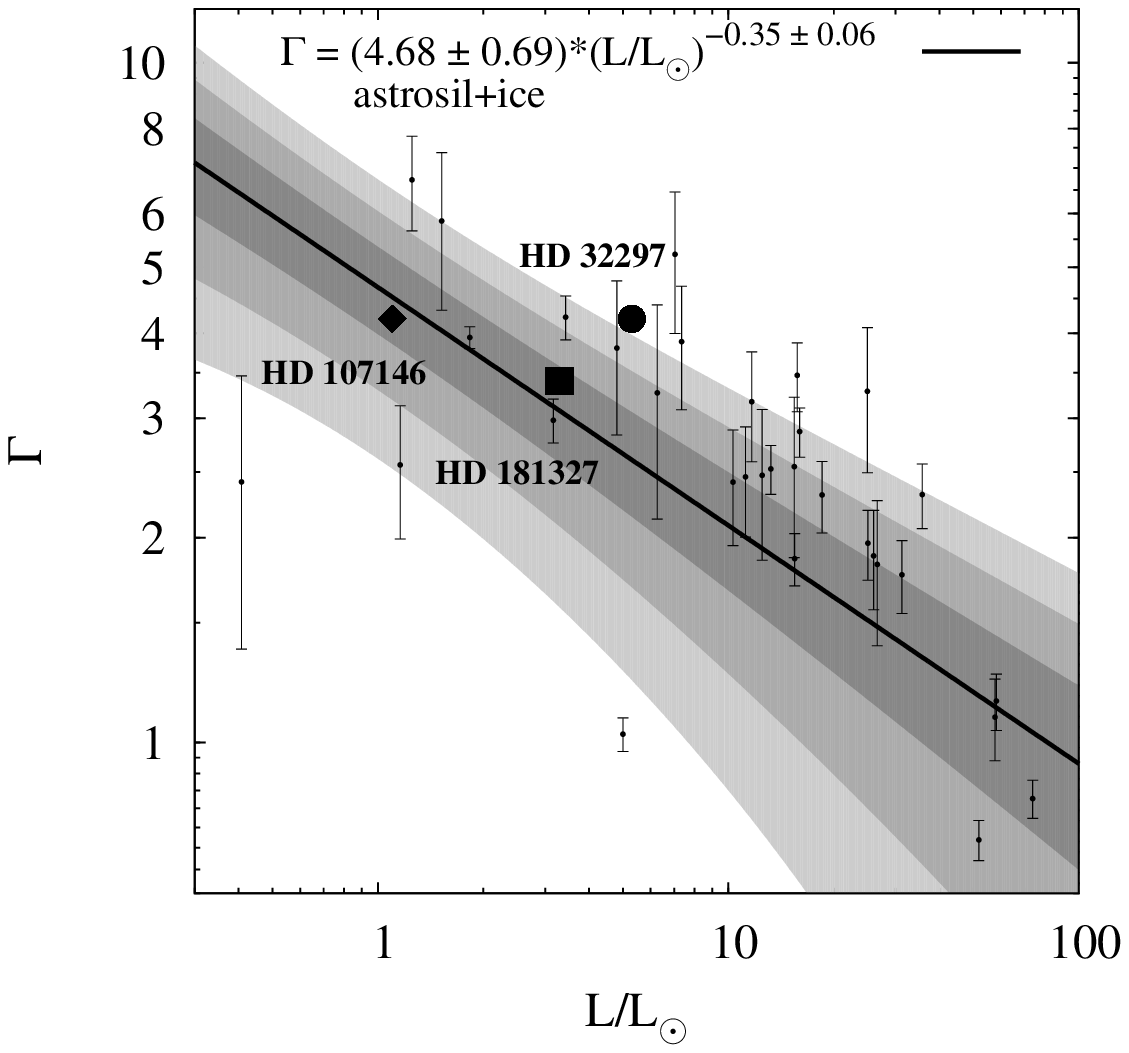}
   \includegraphics[width=0.5\textwidth,angle=0]{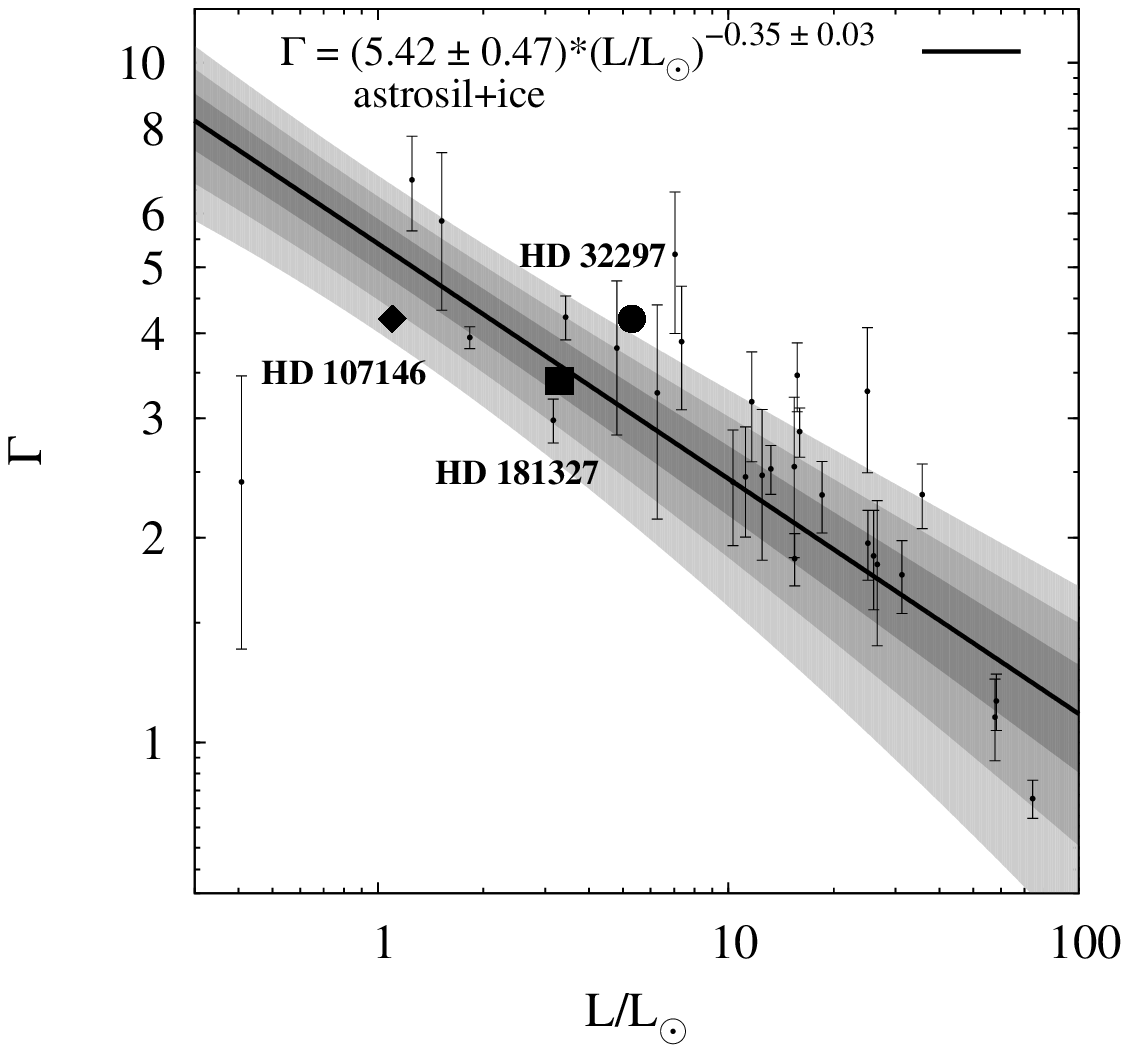}
   \vspace*{-5mm}
   \caption{The predicted ratio of the true disc radius to the blackbody radius, $\Gamma$,
   for the astrosilicate--ice mixture.
   The solid line is the best-fit relation, Eq.~(\ref{Gamma}).
   The areas filled with dark-, medium-, and light-grey are $1\sigma-$, $2\sigma-$, and $3\sigma-$confidence areas for that
   relation, respectively. The symbols show the actual $\Gamma$ values (i.e., observed versus blackbody radius)
   for three selected test discs.
   Unlabeled small symbols with error bars are fitting results for the discs in our sample, for the same mixture.
   These are the same as blue symbols in Fig.~\ref{fig:parm_L}b.
   Top: the entire sample.
   Bottom: three outliers (HD 104860, $\eta$~Crv, and Vega) excluded.
   \label{fig:test_discs}
   }
   \end{figure}

\begin{table}
\caption{
Power-law coefficients for the $\Gamma$ ratio
\label{tab:coeff}
}
\tabcolsep 3pt
  \begin{center}
    \begin{tabular}{lcc}
    \toprule
  Dust composition& $A$& $B$\\
  \midrule
  50\% astrosilicate + 50\% vacuum        & $5.75 \pm 0.66$ & $-0.40 \pm 0.04$\\
  50\% astrosilicate + 50\% ice           & $5.42 \pm 0.47$ & $-0.35 \pm 0.03$\\
  100\% astrosilicate                     & $8.26 \pm 1.27$ & $-0.55 \pm 0.04$\\
  50\% astrosilicate + 50\% carbon        & $6.44 \pm 0.80$ & $-0.41 \pm 0.04$\\
  100\% \revision{carbon}                 & $6.61 \pm 0.74$ & $-0.45 \pm 0.04$\\
  \bottomrule
  \end{tabular}
  
\end{center}

\end{table}

\begin{table*}
\caption{
Resolved discs used to test the $\Gamma(L)$ relation.
\label{tab:test_discs}
}
\tabcolsep 3pt
  \begin{center}
    \begin{tabular}{rcccllr}
  \toprule
  HD number	&$L/\Ls$	&$\td[K]$	&$\Rbb$[\AU]	&$\Gamma_\mathrm{pred}$		&$R_\mathrm{pred}[\AU]$	&$R_\mathrm{true}$[\AU]\\
  \midrule
  107146	&1.1 		&51		&31		&5.5 (astrosil+vacuum) 		&172 (astrosil+vacuum)	&130\\
  		&  		&		& 		&5.2 (astrosil+ice)		&163 (astrosil+ice)	&   \\
  		&  		&		& 		&7.8 (astrosil)			&243 (astrosil)		&   \\
  		&  		&		& 		&6.2 (astrosil+carbon)		&192 (astrosil+carbon)	&   \\
  		&  		&		& 		&6.3 (carbon)			&196 (carbon)		&   \\
  \midrule
  181327	&3.3 		&73		&26		&3.6 (astrosil+vacuum)		& 93 (astrosil+vacuum)	& 89\\
  		&  		&		& 		&3.6 (astrosil+ice)		& 93 (astrosil+ice)	&   \\
  		&  		&		& 		&4.3 (astrosil)			&112 (astrosil)		&   \\
  		&  		&		& 		&3.9 (astrosil+carbon)		&103 (astrosil+carbon)	&   \\
  		&  		&		& 		&3.9 (carbon)			&101 (carbon)		&   \\
  \midrule
   32297	&5.3 		&83		&26		&3.0 (astrosil+vacuum)	 	& 77 (astrosil+vacuum)	&110\\
  		&  		&		& 		&3.0 (astrosil+ice)		& 79 (astrosil+ice)	&   \\
  		&  		&		& 		&3.3 (astrosil)			& 89 (astrosil)		&   \\
  		&  		&		& 		&3.2 (astrosil+carbon)		& 84 (astrosil+carbon)	&   \\
  		&  		&		& 		&3.1 (carbon)			& 92 (carbon)		&   \\
  \bottomrule
  \end{tabular}
  
\end{center}

\end{table*}

Relation (\ref{Gamma}) can be applied to estimate the radius of an unresolved debris disc.
This is particularly useful,
because most of the known debris discs are unresolved.
The recipe is as follows.
First, one takes an SED and fits one or another model to it
\citep[e.g., modified blackbody emission,][]{backman-paresce-1993}
to derive the dust temperature $\td$. This temperature can often be found
in the original papers reporting on the newly discovered excess sources, since
such papers usually perform an SED fitting for these sources as well.
Alternatively, $\td$ can be estimated directly from the wavelength $\lambda_\mathrm{max}$
at which the excess emission peaks: $\td = 5100 \K ( 1 \mum / \lambda_\mathrm{max})$.
Second, one uses $\td$ and the stellar bolometric luminosity to compute the blackbody radius of the disc:
\be
 \Rbb = \left( 278 \K \over \td \right)^2 
        \left( L \over \Ls \right)^{1/2} \AU.
\ee
Third, to get the ``true'' disc radius estimate, one multiplies $\Rbb$ by $\Gamma$ given by
Eq.~(\ref{Gamma}) with coefficients $A$ and $B$ taken from Table~\ref{tab:coeff} for the dust
composition that is thought to be the most appropriate. The choice of the composition can be
guided by the SED fitting itself. Indeed, SED fitting procedures
usually include testing various possible compositions, reporting those that deliver the smallest
$\chi^2$ values.

To illustrate how the method works and to test how reliable it is, we can estimate
the radii of a few resolved discs that are not in our sample, i.e., were not used to derive
the $\Gamma(L)$ relations in this paper. The results can then be comfortably compared to the
actual radii, as measured from the resolved images. Since our sample
only includes discs resolved by the {\em Herschel}/PACS instrument
in thermal emission, the suitable candidates for our test can easily be found among the discs
resolved in scattered light. We arbitrarily chose three such discs, namely those around
a G2V star HD~107146 \citep[taking the data from][]{ardila-et-al-2004,williams-et-al-2004,ertel-et-al-2011},
an F5V star HD~181327 \citep{schneider-et-al-2006,lebreton-et-al-2012},
and an A7 star HD~32297 \citep{kalas-2005,schneider-et-al-2005,donaldson-et-al-2013}.
We then applied the procedure described above, separately for each of the five dust compositions.

The results are listed in Table~\ref{tab:test_discs},
and we deem them very reasonable.
An exception is the radius estimates of the disc of the solar-type star
HD~107146 obtained by assuming strongly absorbing compositions (astrosilicate, carbon, and their mixture).
Both for HD~107146 and HD~181327, a better prediction for the disc radius
is made by assuming a mixture of astrosilicate with ice or vacuum,
which is consistent with \citet{lebreton-et-al-2012} who favour such mixtures
over pure astrosilicate.
Conversely, in the case of HD~32297, the discrepancy between the predicted and the 
observed radius is the smallest for more strongly absorbing materials.
A study over larger samples would be useful to check whether this is a chance effect or an
indication of higher abundances of volatiles and/or higher porosity of dust around lower-luminosity stars.

Figure~\ref{fig:test_discs} puts all three test objects onto the $\Gamma(L)$ plot
for the astrosilicate-ice mixture,
showing the confidence areas for
the $\Gamma(L)$ fit derived above. It demonstrates that the worst of our three cases, HD~32297, is a
$\approx 3\sigma$-outlier. However, black symbols show that there are a number of $3\sigma$-outliers in our sample as well,
so that the case of HD~32297 is nothing extraordinary. Therefore, this example gives us a good
grasp on the accuracy of the method. Underestimating or overestimating the true radius by up to a factor
of two is possible. On any account, applying our method ensures a much better estimate of the true radius
($89\AU$--$130\AU$)
than just taking the blackbody values ($26\AU$--$31\AU$ for our three discs).

\section{Explaining the trend: The role of the surface energy constraint}
\label{sec:K&K}

We now turn to possible explanations for the size-luminosity trend analysed in preceding sections.
One possibility is that smaller collisional fragments around stars of lower luminosity
may not be produced at all.
Recently, \citet{krijt-kama-2014} pointed out that the minimum size of the collisional fragments
created in the collisional cascade should be limited by the available impact energy.
Denoting the fraction of kinetic energy that is spent to create the surfaces of
collisional fragments by $\eta$, and the surface energy of a unit surface by $\gamma$,
the minimum fragment radius is given by their~Eq.~(7):
\be
  x
  \equiv
  {\smin \over \sblow}
  = 48 A
  \left(0.01 \over f \right)^2 ,
\label{K&K}
\ee
where \change{Typo in the equation fixed}
\be
  A
  \equiv
  \left(\rdisc \over 100\AU \right)
  \left(\Ls \over L \right)
  \left(0.01 \over \eta \right)
  \left(\gamma \over 100\erg\cm^{-2} \right) 
\label{A}
\ee
\revision{and $f$ is the mean relative velocity of colliders in the units of Keplerian velocity at
the distance $\rdisc$ from the star.
Following \citet{krijt-kama-2014}, $f$ is set equal to
the average eccentricity of the dust parent planetesimals, $\emax$.}

\revision{Figure~\ref{fig:K&K} (top) depicts the size ratio as a function of stellar luminosity
for a subsample of 14 discs with fractional luminosities higher than the median value.
As shown in section~\ref{sec:subsamples},
this subsample may be more reliable than the entire one.
The discs with high fractional luminosity reveal a stronger correlation of the $\sratio$ ratio
with the stellar luminosity, which allows an easier comparison with the models.}

\revision{The $\sratio$ ratios as functions of $L/\Ls$ given by Eq.~(\ref{K&K}) are}
overplotted in Fig.~\ref{fig:K&K} (top) with dashed lines for several
values of $\emax$.
In that calculation, we assumed
$\eta = 0.01$ and 
$\gamma = 100\erg\cm^{-2}$,
with the caveat that these parameters are very uncertain.
\revision{The regions under the dashed lines correspond to particles that cannot be created
in collisions, because the impact energy would not be sufficient to create their surfaces.
Higher eccentricities allow smaller grains to be produced.
For $\emax \approx 0.02$--$0.03$, 
the regions excluded by this constraint,
together with the regions excluded by the blowout limit,
match those areas where no data points are present.}

   \begin{figure}
   \centering
   \includegraphics[width=0.5\textwidth,angle=0]{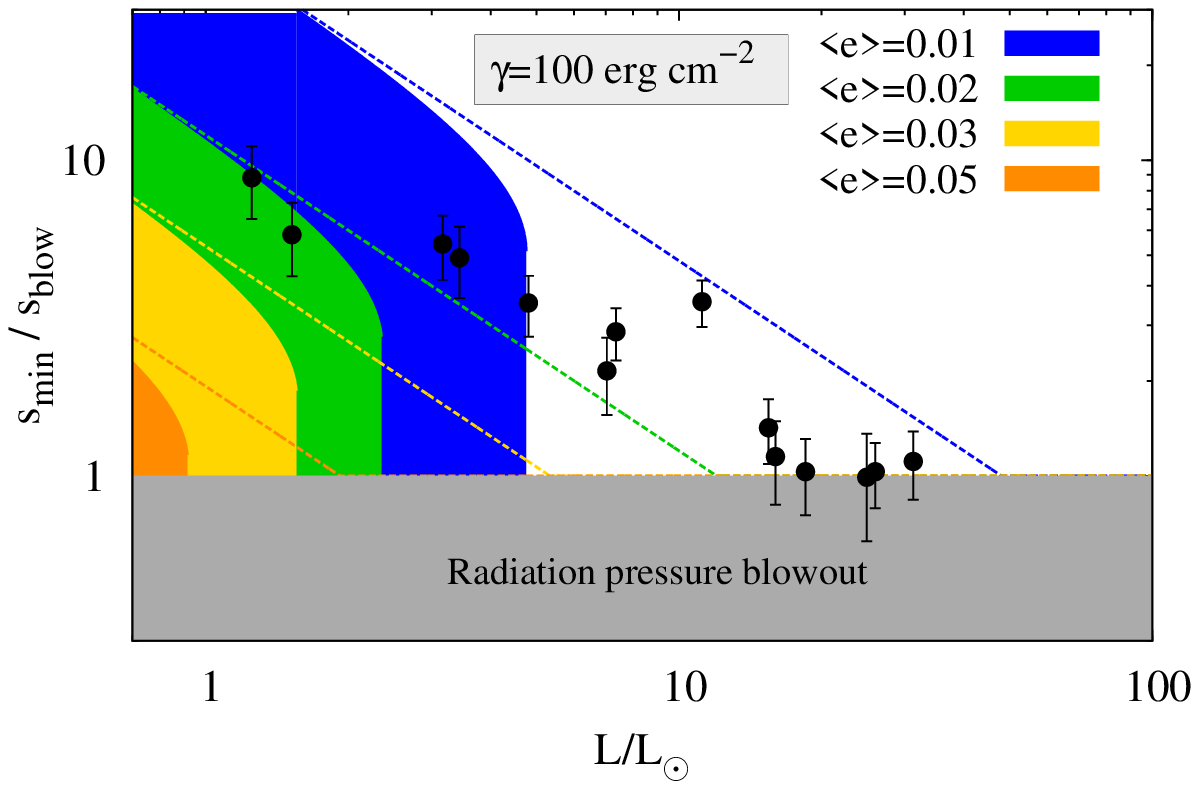}
   \includegraphics[width=0.5\textwidth,angle=0]{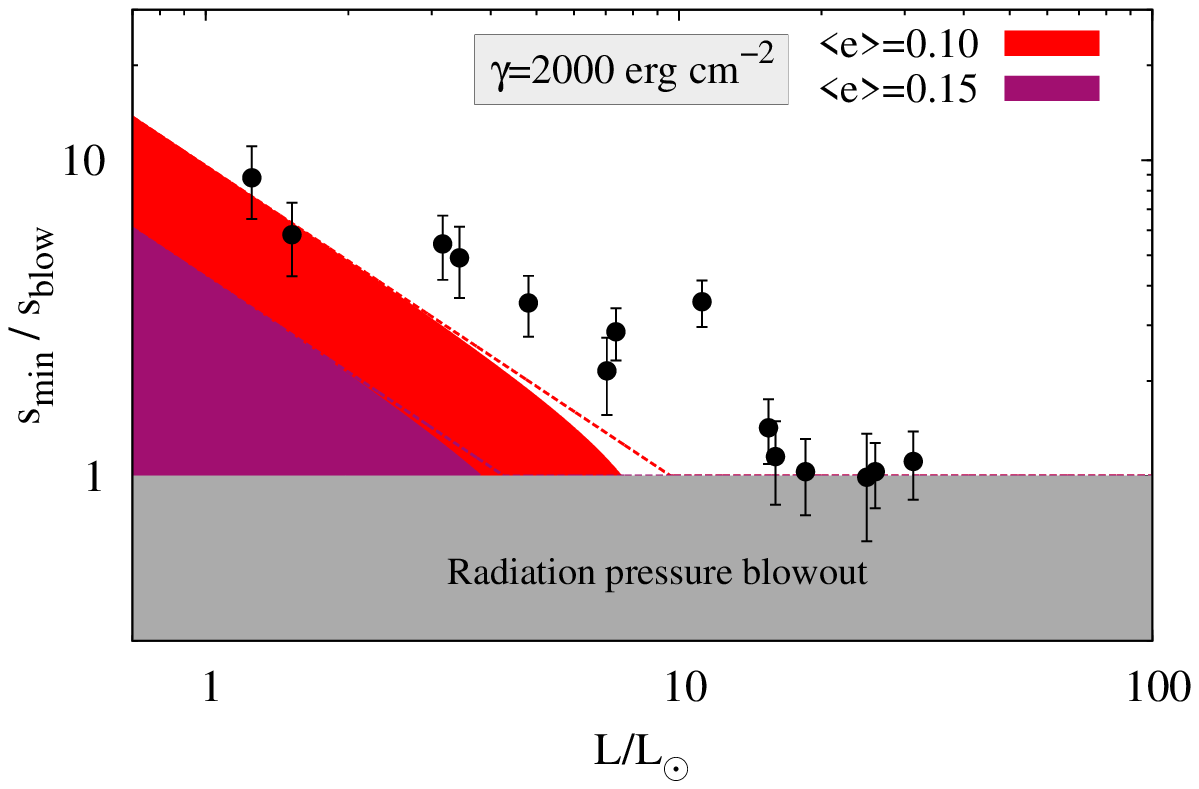}
   \vspace*{-5mm}
   \caption{
   \change{Figure reworked.} 
   The $\sratio$ ratio as a function of stellar luminosity, assuming pure astrosilicate
   grains. Symbols with error bars: \revision{discs of our sample with $\Fd > 7.26 \times 10^{-5}$}.
   Gray-shaded area: region of blowout grains.
   Dashed lines: original model by \citet{krijt-kama-2014}, Eq.~(\ref{K&K}).
   Filled areas: improved model, Eq.~(\ref{K&K improved}).
   \revision{Grains below the dashed lines and those in filled areas are excluded by these models,
   because they should not be produced.}
   Different line and filling colours correspond to different degrees of the dynamical excitation, $\emax$.
   Top:  $\gamma = 100\erg\cm^{-2}$, symbols represent the full sample;
   bottom:  $\gamma = 2000\erg\cm^{-2}$.
   \label{fig:K&K}
   }
   \end{figure}
   
\revision{As noted by \citet{krijt-kama-2014},
there is an obvious possibility to further improve the model
by taking into account that radiation pressure should excite eccentricities
of dust grain orbits to values higher than those of their parent planetesimals.
Accordingly,} it would be more accurate
to attribute $f$ to the average eccentricity of smaller solids~-- namely those, 
which are ``immediate parents'' of the collisional fragments in question.
We now assume that this role is played by grains with radius $b \smin$,
where $b>1$ is a numerical factor.
These grains inherit the dynamical excitation of planetesimals, but
their eccentricities are further increased by radiation pressure.
We can adopt
\be
  f \approx \sqrt{\emax^2 + \left(\beta \over 1-\beta\right)^2},
\ee
where $\beta$ is given by Eq.~(\ref{beta}).
Setting for simplicity $\Qpr$ to unity, which is also an assumption behind
Eqs.~(\ref{K&K})--(\ref{A}), we write $\beta = 1/(2 b x)$.
Equation~(\ref{K&K}) generalises to
\be
  x
  = 48 A
  \left(0.01 \over \emax \right)^2 
  \left[
        1
        +
        {1 \over \emax^2 (2bx-1)^2}
  \right]^{-1} .
\label{K&K improved}
\ee
This is a cubic equation for $x$, the largest root of which is the
$\sratio$ ratio we are seeking.  
The original equation (\ref{K&K}) can be obtained from
(\ref{K&K improved}) as a limit $b \rightarrow \infty$.

\revision{Since Eq.~(\ref{K&K improved}) is a cubic equation, it has two additional
smaller roots.
It is interesting to find out whether they also have any physical meaning. It turns out
they do. The fragments with $\sratio$ ratio lying between these two smaller roots
correspond to the case where parent grains have sufficiently
large radiation pressure-induced eccentricities to allow production of these
fragments as well. This can most easily be understood by looking at another limiting case,
$\emax \rightarrow 0$. In that case, Eq.~(\ref{K&K improved}) transforms to a quadratic
\be
  x
  = 4.8 \times 10^{-3} A
  \left( 2bx-1 \right)^2 .
\ee
For the size ratios between the two roots of this equation, the impact energies,
owing to radiation pressure-induced eccentricities, are large enough to allow creation
of fragments. However, for the parameter ranges considered here both roots of that equation
are smaller than unity, i.e., correspond to $\smin < \sblow$.
Thus they are not physically relevant and are not considered further.
}

The regions excluded by Eq.~(\ref{K&K improved})
(i.e., the ratios $\sratio$ less than $x$, where
$x$ is the \revision{largest root of} Eq.~\ref{K&K improved})
are shown in Fig.~\ref{fig:K&K} (top) 
as filled areas, for the same set of eccentricities $\emax$.
In that calculation, we \revision{set $b = 10$, thus assuming that collisional fragments are on the average
ten times smaller than the target.}
\citep[See][for justification of this choice and additional references.]{krivov-et-al-2005}

\revision{Looking at the shape of the filled regions shown in Fig.~\ref{fig:K&K} (top),
the refined model may seem to reproduce the trend in the data points more poorly than the original model.
However, a closer similarity between the model constraints and the data can be achieved
by varying the coefficient $A$ in Eq.~(\ref{K&K improved}). 
This is demonstrated by Fig.~\ref{fig:K&K} (bottom) where we
increased the coefficient $A$ (see Eq.~\ref{A}) by a factor of 20.
The latter may correspond,
for instance, to a 20 times higher surface energy per unit surface area, $\gamma$
(still not unrealistic, e.g., for icy material) or to a 20 times lower energy fraction
that goes to the surface creation, $\eta$ (not unrealistic either).
With this choice, a reasonable match to the trend seen in the data is achieved for
somewhat higher values of $\emax$.
Indeed, one sees} that for $\emax \approx 0.10$, the region in 
Fig.~\ref{fig:K&K} (bottom) excluded by our model could masquerade as a decrease
in $\sratio$ seen in the data.

\section{Explaining the trend: The role of the stirring level}

In this section, we check another possible explanation for the trend in grain sizes with the
stellar luminosity. The mechanism that we will address here is \revision{dissimilar, albeit not
completely unrelated, to} the microphysical one discussed in the previous section.
\revision{It is associated with} the balance between
the production and loss rate of small grains, controlled by the stirring level of larger bodies.

\subsection{Idea}

\citet{thebault-wu-2008} predicted the grain size distribution
to depend on the degree of dynamical excitation of the dust-producing 
planetesimals, $\emax$.
If the planetesimals have a low dynamical excitation which, however, is still high enough for
collisions to be mostly destructive,
then the low collision velocities between large grains, that are not susceptible
to radiation pressure, would decrease the rate at which small
grains are produced. However, the destruction rate of these small grains is set by
eccentricities induced by radiation pressure and remains the same.
This should result in a dearth of small dust. The maximum of the size
distribution, i.e. the ratio $\sratio$, would shift to larger values. 
This is easy to quantify.
Following the explanation above, the size distribution should peak at grain sizes
$\smin$, for which the radiation pressure-induced eccentricity, $\beta/(1-\beta)$, equals
the eccentricity inherited from the planetesimals, $\emax$.
Since $\beta \approx 0.5(\sblow/\smin)$ (assuming $\Qpr = 1$), this gives
\be
 \sratio \approx \left( \emax^{-1} + 1 \right)/2 .
\label{th&wu}
\ee

However, Eq.~(\ref{th&wu}) is just a rough estimate.
Obviously, the effect has to be confirmed by collisional simulations that 
include a more realistic treatment of collisional and radiation pressure forces.
Such a study is still missing,
since the paper by \citet{thebault-wu-2008} was confined to discs of 
A-stars, and the size distribution was computed with a collisional code 
that did not include cratering collisions, rebounds and sticking.

\subsection{ACE runs}

To quantify the effect more accurately, we performed four runs of our ACE code 
\citep{krivov-et-al-2013},
probing two central stars (A2V with $17.4\Ls$ and G2V with $1\Ls$)
and two typical stirring levels (average planetesimals' eccentricity of
0.1 and 0.01, average inclination according to the energy equipartition).
The simulations included stellar gravity, direct radiation pressure and
Poynting-Robertson drag, and a wealth of possible collisional outcomes
(disruptive, cratering, rebounding, and sticking collisions).
In all of the cases, the initial disc mass was taken to be
$30 \me$ (in the bodies of up to $100\km$ radius), in order to arrive at
the typical dust fractional luminosity level in our sample.
The disc radius was set to $100\pm 10 \AU$.
We also made a number of other standard assumptions.
In particular, we assumed compact astrosilicate from 
\citet{draine-2003} as a material composition,
the critical fragmentation energy from
\citet{benz-asphaug-1999}, etc.
Each disc was evolved until a quasi-steady state, as defined 
in \citet{loehne-et-al-2007}, was reached.

\subsection{Results}

The resulting size distributions are shown in Fig.~\ref{fig:ACE_sizes}.
They confirm the effect of the maximum in the size distribution shifting towards larger values
for discs with lower dynamical excitation. However, the size distributions simulated with ACE
cannot be closely approximated by power laws with a sharp lowest cutoff.
In the case of G2-star discs ($\sblow = 0.46\mum$),
the position of the maximum in the size distribution,
$2.6\mum$ for $\emax=0.1$ or
$10\mum$ for $\emax=0.01$,
can still be taken as $\smin$.
In the discs around an A2-star, the size distribution develops a ``plateau''
between $\sblow = 3.0\mum$ and a shallow maximum
at $12\mum$ (for $\emax=0.1$) or $50\mum$ (for $\emax=0.01$).
The different-sized grains in these ranges contribute to the cross section almost equally.
In that case, we take a geometric mean between the two ends of the plateau as a proxy
for $\smin$.

   \begin{figure}
   \centering
   \includegraphics[width=0.5\textwidth,angle=0]{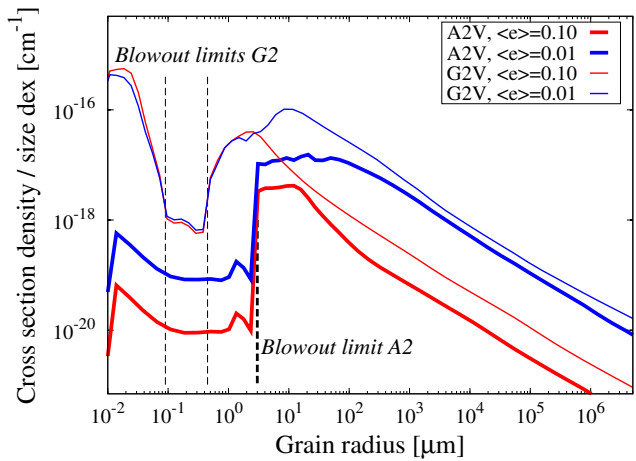}
   \vspace*{-2mm}
   \caption{Simulated size distributions in the fiducial discs around A2-stars (thick lines)
   and G2-stars (thin lines) with higher (red) and lower (blue) level of stirring.
   Plotted is the size distribution
   in the parent ring, assumed to be located at $100\AU$.
   Vertical dashed lines mark the radiation pressure blowout limit of A2-stars (thick) and G2-stars (thin).
   Note that two blowout values exist for the G2-stars.
   The grains in unbound obrits are those between these two.
   \label{fig:ACE_sizes}
   }
   \end{figure}
   
   \begin{figure}
   \centering
   \includegraphics[width=0.5\textwidth,angle=0]{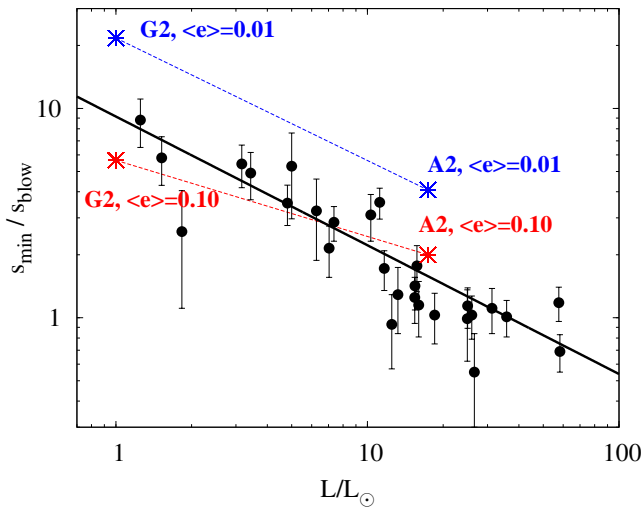}\\
   \vspace*{-5mm}
   \caption{
   Symbols with error bars represent the discs of our sample
   (assuming pure astrosilicate)
   and the thick black line is the best-fit through these data.
   Asterisks mark approximate values of $\sratio$ extracted from ACE simulations, as
   described in the text.
   The pairs of asterisks corresponding to the same dynamical excitation, $\emax$, are connected with
   dashed lines for illustrative purposes.
   \label{fig:ACE_stirring}
   }
   \end{figure}

Figure~\ref{fig:ACE_sizes} also shows that for the G2V stars, two blowout limits exist.
The ``sub-blowout grains'' of sizes smaller than $\la 0.1\mum$ are also is bound orbits
and make a large contribution to the total cross section of dust. However, their absorption
efficiency is much lower than that of the large bound grains.
Accordingly, their contribution to the observed emission is minor, at $\sim 10$\% level at all wavelengths.
For this reason, the sub-blowout grains are not directly relevant to the above discussion of the lower cutoff
in the size distribution. Nevertheless, these grains still implicitly affect the size distribution of
large bound grains, since collisions with them ``erode'' the population of particles with sizes just 
above the main blowout size.

Figure~\ref{fig:ACE_stirring} 
compares
the resulting $\sratio$ ratios found with the ACE runs
with our sample.
As expected, lower $\emax$ lead to larger $\sratio$,
and $\smin$ are roughly consistent with Eq.~(\ref{th&wu}).
What was not really expected prior to our modelling, however, is that the
lines of constant dynamical excitation are tilted.
The collisional modelling shows that for stars of earlier spectral types,
$\smin$ becomes closer to $\sblow$ even if $\emax = \mathrm{const}$~---
an effect that was not previously reported in the literature.
Coincidentally or not, this is exactly the qualitative trend we found in the observational data.

However, the observed trend is stronger, and is
not consistent with a constant $\emax$ assumption for stars across the full
luminosity range.
Instead, the data
seem to be best reproduced by assuming $\emax$ between 0.01 and 0.1 for G-stars and
$\emax \ga 0.1$ for A-stars.
This impression is supported by the detailed collisional modelling of several individual objects
done previously.
For instance, \citet{loehne-et-al-2011} and \citet{schueppler-et-al-2014} find $\emax$ to be at the level
of a few per cent for the discs of a G0V-star HD~207129 and a K2V-star HIP~17439, respectively,
whereas higher values $\emax \sim 0.1$ are favoured for the discs of an A5V-star HR8799
\citep[where it is expected based on the presence
of a substantial outer halo of small grains, see][]{matthews-et-al-2013b}
and an A0V-star Vega 
\citep{mueller-et-al-2009}.
We stress, however, that these claims should not be overinterpreted.
We are only discussing general statistical trends. Individual systems show an appreciable
scatter, and there are certainly discs that do not follow those general trends.

There is one more aspect related to the stirring level.
If the discs around more luminous stars are excited more strongly, they should possess
more pronounced halos. In that case, the disc radii of early-type stars
retrieved from the images may be overestimated, and so, the dust grain sizes
may be underestimated \citep{pawellek-et-al-2014}.
Correcting for this effect would push the data points on the right
of Fig.~\ref{fig:ACE_stirring} up,
flattening the $\sratio$ dependence on stellar luminosity.
However, this would make the excitation level of these discs estimated from the plot
lower than before the correction.
This would imply the weaker halos, pushing the data points of luminous stars back down
and resulting in smaller $\sratio$ at high luminosities and a steeper
dependence. Thus, the $\sratio (L)$ dependence plotted in
Fig.~\ref{fig:ACE_stirring} may be, so to say, ``self-regulating'' and should be
relatively robust with respect to the uncertainties in measuring the disc radius
from the images.

\revision{Another remark is that the strirring models considered here and the surface energy models
addressed in section~\ref{sec:K&K} are not independent.
If the stirring level does depend on stellar luminosity as proposed here,
this will affect the surface energy constraint discussed in section~\ref{sec:K&K},
changing the filled areas shown in Fig.~\ref{fig:K&K}.
The best way to address this would be to combine both effects in a single model.
This can be done in the future by implementing the surface energy constraint directly 
into the kernel of the collisional code that controls the production of fragments
(the ACE simulations presented here did not include the surface energy
constraint).}

Can we expect the discs of more luminous stars to be more strongly stirred on the average than those of
low-luminosity stars?
In principle, yes.
Indeed, it is known that the submillimeter dust masses in protoplanetary discs, the progenitors of debris
discs, are roughly proportional to the masses of their central stars
\citep[see, e.g., Fig.~5 in][and discussion therein]{williams-cieza-2011}.
It is possible that discs of more massive (or more luminous) stars,
being more massive,
were more ``successful'' in
building large planetesimals that may later act as stirrers for debris discs
\citep[the so-called ``self-stirring scenario'', e.g.,][]{wyatt-2008,kennedy-wyatt-2010}.
This is directly supported by planetesimal accretion simulations
\citep[e.g.,][]{kenyon-bromley-2008}.
Alternatively or additionally,
such protoplanetary discs may have formed more giant planets, and/or these planets may
have experienced more vigorous migration or scattering
in the past, which might have also resulted in a higher dynamical excitation of the debris discs
emerged in these systems
\citep[the ``planetary stirring scenario'', e.g.,][]{mustill-wyatt-2009}.
Indeed, observational \citep[e.g.,][]{johnson-et-al-2010,reffert-et-al-2015} and
theoretical work \citep[e.g.,][]{ida-lin-2005,kennedy-kenyon-2008,%
mordasini-et-al-2011b,mordasini-et-al-2012} find that
giant planets are more frequent around more massive stars, and that those
planets are typically more massive.
Further work is required, however, to validate or falsify these possibilities.

\section{Conclusions and discussion}

Analysing a sample  of Herschel-resolved debris discs, \citet{pawellek-et-al-2014} found
the ratio of the minimum (or typical) grain size $\smin$ to the radiation pressure blowout size $\sblow$
to systematically decrease with the increasing luminosity of the central stars.
Here we investigate how robust the trend is and attempt to find possible explanations for it.
Our conclusions are as follows:

\begin{enumerate}
\item
The decrease of $\sratio$ with increasing luminosity of the central stars
persists, no matter which material compositions and porosity of dust grains are assumed.
The trend is gentler for porous grains, but can be completely erased only for 
unrealistically high porosities.

\item
The minimum grain size itself is consistent with being constant, $\smin \approx 5\mum \pm 0.3\mum$,
across the full luminosity range of the sample.

\item
We have tested the subsamples of discs with lower and higher fractional luminosity,
smaller and larger radii, younger and older ages, lower and higher absolute integrated flux,
lower and higher absolute surface brightness, as well as those that are
marginally and well-resolved. In terms of $\sratio(L)$, all of these subsamples are found
to be statistically indistinguishable.
Marginal differences are only visible between the discs with higher and lower fractional luminosity
in our sample.
We argue that these are likely related to SED-fitting aspects rather than reflect any real
physical difference between the dustier and the less dusty discs.

\item
As a by-product of this study, we derive empirical formulae for the ratio $\Gamma$ of the true disc radius
to its blackbody radius, as a function of stellar luminosity. Since the blackbody radius can 
easily be found from
the SED of any excess source, this offers a recipe of how to estimate the true radius of unresolved
debris discs, thus breaking the notoriously known degeneracy between the dust grain sizes and 
the dust location.

\item
The results of our collisional simulations reproduce
the $\sratio$ trend with the stellar luminosity seen in the sample.

\item
Additional effects may also be at work, contributing to the observed trend.
For instance, the surface energy constraint on the size of the smallest collisional fragments identified by
\citet{krijt-kama-2014} may be responsible for the absence of grains with small $\sratio$ ratios
in discs of lower luminosity stars.

\item
It is also possible that the $\sratio$-trend is related to the degree of stirring of the
dust-producing planetesimals.
Indeed,
a better agreement between the data and
the collisional simulations can be achieved by assuming that the discs of earlier-type stars
are more strongly excited (on the average)
than those of later-type stars.
This would imply that protoplanetary discs of more massive young stars are more efficient in
forming big planetesimals or planets that act as stirrers in the debris discs at the subsequent
evolutionary stage.
\end{enumerate}

\revision{Regardless of whether the observed trend is related to microphysics
of collisional dust production or stirring levels of planetesimals,
there is a close relation between these two types of models.
Both involve small grain dynamics being set by radiation pressure.
This emphasizes again that dust grain dynamics is a key ingredient in
debris disc models, without which confronting simulations and observations would be unthinkable.}
 
In principle, it is possible to find alternative explanations for the trend by allowing various parameters of
dust and/or planetesimals to systematically vary with the type of the central stars. These may include, for instance,
the chemical composition, degree of porosity, or tensile strength. However, such an assumption is 
not substantiated by any pieces of observational evidence. Nor is it supported by theoretical 
expectations.
For example, there is no compelling evidence for a systematic difference in the chemical 
composition in the cold parts of the discs
of Herbig Ae/Be stars and those of T~Tau stars, which are progenitors of debris discs around early and later-type stars, respectively
\citep[e.g.,][]{dutrey-et-al-2013}.

\section*{Acknowledgements}

We thank Torsten L\"ohne for numerous stimulating discussions
and the referee for enlightening and constructive comments.
Support by the DFG through
\revision{grants Kr~2164/10-1, Kr~2164/13-1, and Kr~2164/15-1}
is acknowledged.




\newcommand{\AAp}      {\hbox{A{\&}A}}
\newcommand{\AApR}     {Astron. Astrophys. Rev.}
\newcommand{\AApS}    {AApS}
\newcommand{\AApSS}    {AApSS}
\newcommand{\AApT}     {Astron. Astrophys. Trans.}
\newcommand{\AdvSR}    {Adv. Space Res.}
\newcommand{\AJ}       {AJ}
\newcommand{\AN}       {AN}
\newcommand{\AO}       {App. Optics}
\newcommand{\ApJ}      {ApJ}
\newcommand{\ApJL}     {ApJL}
\newcommand{\ApJS}     {ApJ Suppl.}
\newcommand{\ApSS}     {Astrophys. Space Sci.}
\newcommand{\ARAA}     {ARA{\&}A}
\newcommand{\ARevEPS}  {Ann. Rev. Earth Planet. Sci.}
\newcommand{\BAAS}     {BAAS}
\newcommand{\CelMech}  {Celest. Mech. Dynam. Astron.}
\newcommand{\EMP}      {Earth, Moon and Planets}
\newcommand{\EPS}      {Earth, Planets and Space}
\newcommand{\GRL}      {Geophys. Res. Lett.}
\newcommand{\JGR}      {J. Geophys. Res.}
\newcommand{\JOSAA}    {J. Opt. Soc. Am. A}
\newcommand{\MemSAI}   {Mem. Societa Astronomica Italiana}
\newcommand{\MNRAS}    {\hbox{MNRAS}}
\newcommand{\PASJ}     {PASJ}
\newcommand{\PASP}     {PASP}
\newcommand{\PSS}      {Planet. Space Sci.}
\newcommand{\RAA}      {Research in Astron. Astrophys.}
\newcommand{\SolPhys}  {Sol. Phys.}
\newcommand{\SolSysRes}{Sol. Sys. Res.}
\newcommand{\SSR}      {Space Sci. Rev.}


\input paper.bbl.std



%
%


\bsp	
\label{lastpage}
\end{document}